 \documentclass[10pt,preprint]{aastex}  
 \usepackage{natbib}
\def\ang{\AA}
\def\arcsec{\hbox{$^{\prime\prime}$}}

\def\gapprox{\lower.4ex\hbox{$\;\buildrel >\over{\scriptstyle\sim}\;$}}
\def\lapprox{\lower.4ex\hbox{$\;\buildrel <\over{\scriptstyle\sim}\;$}}

\shortauthors{ASCHWANDEN ET AL. 2014}
\shorttitle{Global Energetics of Solar Flares. II.}

\begin{document}

\title{         Global Energetics of Solar Flares: 
		II. Thermal Energies }

\author{        Markus J. Aschwanden$^1$ and Paul Boerner$^1$}

\affil{		$^1)$ Lockheed Martin, 
		Solar and Astrophysics Laboratory, 
                Org. A021S, Bldg.~252, 3251 Hanover St.,
                Palo Alto, CA 94304, USA;
                e-mail: aschwanden@lmsal.com }

\author{        Daniel Ryan$^2$}


\affil{		$^2)$ Royal Observatory of Belgium,
		Solar-Terrestrial Centre for Excellence,
		Avenue Circulaire 3,
		1180 Uccle, Brussels, Belgium;
 		e-mail: ryand5@tcd.ie }

\author{	Amir Caspi$^3$}

\affil{		$^3)$ Planetary Science Directorate,
		Southwest Research Institute,
		Boulder, CO 80302, USA;
		e-mail: amir.caspi@swri.org }

\author{	James M. McTiernan$^4$}

\affil{		$^4)$ Space Sciences Laboratory,
		University of California, 
		Berkeley, CA 94720, USA;
		e-mail: jimm@ssl.berkeley.edu } 
\and

\author{	Harry P. Warren$^5$}

\affil{		$^5)$ Space Science Division,
		Naval Research Laboratory,
		Washington, DC 20375, USA;
		e-mail: harry.warren@nrl.navy.mil }

\begin{abstract}
We present the second part of a project on the global energetics 
of solar flares and coronal mass ejections (CMEs) that includes about
400 M- and X-class flares observed with the {\sl Atmospheric Imaging
Assembly (AIA)} onboard the 
{\sl Solar Dynamics Observatory (SDO)} during the first 3.5 years 
of its mission. In this Paper II we compute the differential emission 
measure (DEM) distribution functions and associated multi-thermal energies,
using a spatially-synthesized Gaussian DEM forward-fitting method. 
The multi-thermal DEM function yields a significantly higher (by an 
average factor of $\approx 14$), but more comprehensive 
(multi-)thermal energy than an isothermal energy estimate from 
the same AIA data.
We find a statistical energy ratio of $E_{th}/E_{diss} \approx 2\%-40\%$
between the multi-thermal energy $E_{th}$ and the 
magnetically dissipated energy $E_{diss}$, 
which is an order of magnitude higher than the estimates
of Emslie et al.~2012. For the analyzed set of M and X-class flares 
we find the following physical parameter ranges:
$L=10^{8.2}-10^{9.7}$ cm for the length scale of the flare areas,
$T_p=10^{5.7}-10^{7.4}$ K for the DEM peak temperature,
$T_w=10^{6.8}-10^{7.6}$ K for the emission measure-weighted temperature,
$n_p=10^{10.3}-10^{11.8}$ cm$^{-3}$ for the average electron density,
$EM_p=10^{47.3}-10^{50.3}$ cm$^{-3}$ for the DEM peak emission measure,
and $E_{th}=10^{26.8}-10^{32.0}$ erg for the multi-thermal energies.
The deduced multi-thermal energies are consistent 
with the RTV scaling law $E_{th,RTV} = 7.3 \times 10^{-10} \ 
T_p^3 L_p^2$, which predicts extremal values of 
$E_{th,max} \approx 1.5 \times 10^{33}$ erg for the largest flare 
and $E_{th,min} \approx 1 \times 10^{24}$ erg for the smallest 
coronal nanoflare.
The size distributions of the spatial parameters exhibit powerlaw tails that
are consistent with the predictions of the fractal-diffusive
self-organized criticality model combined with the RTV scaling law.
\end{abstract}


\section{		    INTRODUCTION			}

While we measured the magnetic energy that is dissipated in large solar flares 
in Paper I (Aschwanden, Xu, and Jing 2014a), the goal of this paper II
is the determination of thermal energies of the heated flare plasma,
in order to study the energy partition of the input (magnetic) energy
into various output (thermal and other) energies. A crucial test is
the thermal to the magnetic energy ratio, which is expected to be
less than unity for magnetic energy release processes (such as magnetic
reconnection). Ratios in excess of unity would indicate either inaccurate
energy measurements of either magnetic or thermal energies, or would
challenge standard flare scenarios where the source of dissipated 
energy is entirely of magnetic origin. In the standard CSHKP magnetic
reconnection model (Carmichael 1964; Sturrock 1966; Hirayama 1974;
Kopp and Pneuman 1976), magnetic reconnection drives the nonlinear
dissipation of magnetic energy, which is then converted partially
into particle acceleration and (precipitation-driven and conduction-driven)
flare plasma heating, for which the thermal energy is naturally expected
to be a fraction of the total dissipated magnetic energy only.
Thus, statistical measurements of the thermal to magnetic energy ratio
provide crucial tests for theoretical flare scenarios as well as on
the accuracy of observational flare energy measurement methods.

The problematics of determining magnetic energies has been discussed
extensively in Paper I. There are three forms of magnetic energies:
the potential energy, the free energy (or excess of nonpotential
over potential energy), and the dissipated energy, which corresponds
to the negative change of free energy during a flare event. Therefore,
the measurement of dissipated magnetic energies requires methods that
accurately can detect deviations from the potential magnetic field,
which are difficult to achieve, as a quantitative comparison of 
12 nonlinear force-free field (NLFFF) extrapolation methods of the
photospheric magnetic field demonstrated (DeRosa et al.~2009).  
Alternative NLFFF methods that use the geometry of (automatically traced) 
coronal loops as constraints appear to be more promising for this task
(Paper I). There exists
only one study that attempts to compare dissipated magnetic energies
with thermal energies in a set of (large eruptive) flare events
(Emslie et al.~2012), but the dissipated magnetic energy could not be
determined in that study and instead was estimated to amount to 
$E_{diss}/E_p \approx 30\%$ of the potential energy, leading to a 
rather small thermal/magnetic energy ratio, in the order of $E_{th}/E_{diss} 
\approx 0.2\%-1\%$. Since the ratio of the dissipated magnetic energy $E_{diss}$
to the potential magnetic energy $E_p$ has been found to have a substantially smaller
value in Paper I, in the range of $E_{diss}/E_p \approx 1\%-25\%$
for a representative set of M- and X-class flares, we suspect that the 
thermal/magnetic energy ratio is systematically underestimated
in the study of Emslie et al.~(2012). As a consequence, we will see in the 
present study that the thermal/magnetic energy ratio in large solar 
flares is indeed significantly higher than previously inferred in
Emslie et al.~(2012).

Even if we have an accurate method to determine the dissipated magnetic
energy in solar flares, there are also large uncertainties in the determination
of the thermal energy due to the inhomogeneity and multi-thermal nature of the
solar flare plasma. In principle, an accurate measure of the multi-thermal
energy could be determined if the full 3D distribution of electron
temperatures $T_e(x,y,z)$ and electron densities $n_e(x,y,z)$ are known, 
such as produced in 3D magneto-hydrodynamic (MHD) simulations
(e.g., Testa et al.~2012). In practice, we have only 2D images in multiple
wavelengths available to determine the thermal energy. While the lateral
extent (in the [x,y]-plane) of flare-related emission in EUV and soft X-rays
can be accurately measured for instruments with high spatial resolution, 
such as with AIA/SDO, the line-of-sight column depth (in $[z]$-direction)
is subject to geometric models. Moreover, the differential emission measure 
distribution can only be determined as an integral along any line-of-sight,
and thus the thermal inhomogeneity and filling factors along the line-of-sight 
add additional uncertainties. Nevertheless, the presently available
high-resolution and multi-wavelength capabilities of AIA/SDO provide
unprecedented possibilities to model the 3D flare plasma distribution
with much higher fidelity than previous instruments from the 
the {\sl Solar Maximum Mission (SMM)}, the {\sl Solar and Heliospheric
Observatory (SOHO)}, the {\sl Transition Region and Coronal Explorer (TRACE)}, 
and the {\sl Solar-Terrestrial Relationship Observatory (STEREO)} missions. 
It is therefore timely to attempt a statistical
study of magnetic and thermal energies using AIA and HMI data from SDO.

The content of this Paper II includes a description of the data analysis
methods to determine multi-thermal flare energies (Section 2 and Appendix A),
a presentation of observations and results (Section 3 and Tables 1 and 2),
discussions of problems pertinent to the determination of thermal energies
(Section 4), and conclusions about thermal and magnetic flare energies
(Section 5).

\section{ 		DATA ANALYSIS METHODS		 	}

\subsection{		AIA/SDO Temperature Filters		}

The temperature and density analysis carried out here uses EUV images from
the {\sl Atmospheric Imaging Assembly (AIA)} (Lemen et al.~2012;
Boerner et al.~2012) onboard the {\sl Solar Dynamics Observatory (SDO)} 
spacecraft (Pesnell et al.~2012). AIA contains 10 different
wavelength channels, three in white light and UV, and seven EUV channels,
whereof six wavelengths (94, 131, 171, 193, 211, 335 \ang )
are centered on strong iron lines (Fe {\sc viii}, {\sc ix}, {\sc xii},
{\sc xiv}, {\sc xvi}, {\sc xviii}), covering the coronal range from
$T\approx 0.6$ MK to $\gapprox 16$ MK. The 304 \ang\ (He {\sl II})
filter was not used because it is mostly sensitive to chromospheric
temperatures of $T_e \approx 10^{4.7}$, which is outside of the range 
of interest for the flare study here. The calibration of the response 
functions has changed somewhat over time. Early on in the
mission, the response of the 94 and 131 \ang\ channels was underestimated
(see Fig.~10 in Aschwanden and Boerner 2011). Here we will use the currently 
available calibration, which was updated with improved atomic emissivities 
according to the CHIANTI Version 7 code, distributed in the {\sl Solar SoftWare
(SSW)} {\sl Interactive Data Language (IDL)} on 2012 February 13. 

\subsection{ Gaussian Differential Emission Measure Distribution Function }

The measurement of the thermal energy $E_{th} = 3 n_e k_B T_e V$ of an
(isothermal)
flare plasma requires the determination of the electron density $n_e$,
the electron temperature $T_e$, and the flare volume $V$. From
multi-wavelength observations it is customary to calculate the
{\sl differential emission measure (DEM)} distribution function,
which can be integrated over the coronal temperature range and yields
a total emission measure $EM = n_e^2 V$, providing a mean electron
density $n_e$ (for unity filling factor) and a mean DEM peak temperature 
$T_e$. The inference of the DEM can be accomplished either by inversion 
of the observed fluxes using the instrumental response functions,
or by forward-fitting of a suitable functional form of a DEM
distribution function. DEM inversion methods are often unstable 
(Craig and Brown 1976; Judge et al.~1997; Testa et al.~2012; 
Aschwanden et al.~2015), while forward-fitting methods are generally 
more robust, but require a suitable parameterization of an analytical 
function that has to satisfy an acceptable goodness-of-fit criterion.
A comparison of 10 DEM inversion and forward-fitting methods
has been conducted in a recent study with simulated DEMs, using AIA, 
the {\sl EUV Variability Experiment (EVE)}, the {\sl Ramaty High
Energy Solar Spectroscopic Imager (RHESSI}, and the {\sl Geostationary
Orbiting Earth Satellite (GOES)} response functions (Aschwanden et al. 2015),
where the performance of recent DEM methods is discussed in more detail.

One of the most robust choices of a DEM function with a minimum of 
free parameters is a single Gaussian (in the logarithm of the
temperature), which has 3 free parameters only and is defined by 
the peak emission measure $EM_p$, the DEM peak temperature $T_p$, 
and the logarithmic temperature width $w_T$, where the DEM 
parameter has the cgs-units of [cm$^{-5}$ K$^{-1}$], 
\begin{equation}
	DEM(T) = n_e^2 {dz \over dT} = EM_p \exp{ \left(
	- {[log(T) - log(T_p)]^2 \over 2 \ w_T^2 }
	\right)} \ ,
\end{equation}
where the line-of-sight emission measure $dEM = n_e^2 dz$ is 
the temperature integral over the Gaussian DEM (in units of
[cm$^{-5}$], 
\begin{equation}
	EM = \int DEM(T) \ dT \ .
\end{equation}
The Gaussian DEM (Eq.~1) can be forward-fitted to the preflare 
background-subtracted observed fluxes $f_{\lambda}$ in multiple 
wavelengths $\lambda$,
\begin{equation}
	f_{\lambda}(t) =
	F_{\lambda}(t) - B_{\lambda}(t) =
	\int DEM(T) \ R_{\lambda}(T) \ dT
	= \sum_{k=1}^{n_T} DEM(T_k)\ R_{\lambda}(T_k)\ \Delta T_k \ ,
\end{equation}
where $F_{\lambda}$ are the observed fluxes (in units of DN/s)
in the wavelengths $\lambda=94, 131, 171, 193, 211, 335$ \ang,
$B_{\lambda}$ are the observed background fluxes,
$f_{\lambda}$ are the background-subtracted fluxes, 
integrated over the entire flare area, $R_\lambda(T)$ is the 
instrumental response function of each wavelength filter 
$\lambda$ (in units of [DN s$^{-1}$ cm$^5$] per pixel),
and the temperature integration is using discretized 
temperature intervals $\Delta T_k$, which generally are chosen
to be equidistant bins of the logarithmic temperature range.  

In our DEM forward-fitting algorithm we use a temperature range of 
$T_p=0.5-30$ MK that is subdivided equi-distantly into 36 logarithmic
temperature bins $\Delta T_k$, and a Gaussian temperature width range with 
10 values in the range of $w_T=0.1-1.0$. At the same time, the DEM 
peak emission measure value $EM_p$ is evaluated from the median ratio of
the observed to the model (background-subtracted) fluxes,
\begin{equation}
	EM_p = EM_0 \ \left[ 
	{\sum_{\lambda} f_{\lambda}^{obs}  \over
	 \sum_{\lambda} f_{\lambda}^{fit}} \right] \ ,
\end{equation}
where $EM_0=1$ cm$^{-5}$ K$^{-1}$ is the unity emission measure.
The best-fitting values of the peak emission measure $EM_p$,
the peak temperature $T_p$ and temperature width $w_T$ are found 
by a global search in the 2-parameter space $[T, w_T]$ 
and by adjustment of the peak emission measure value $EM_p$.
The best-fit solution is then evaluated by the goodness-of-fit
criterion (e.g., Bevington and Robinson 1992),
\begin{equation}
	\chi(t) = \left[
	{1 \over n_{free}}
	\sum_{\lambda=1}^{n_{\lambda}} 
	{(f_{\lambda}^{fit}(t) - f_{\lambda}^{obs}(t))^2
	\over \sigma_{\lambda}^2(t)} \right]^{1/2} \ ,
\end{equation}
where $f_{\lambda}^{obs}$ are the 6 observed flux values,
$f_{\lambda}^{fit}$ are the flux values of the fitted
Gaussian DEM (Eq.~1), $\sigma_\lambda$ are the estimated
uncertainties, $n_{free}=n_{\lambda}-n_{par}$ is the number 
of degrees of freedom, which is $n_{free}=3$ for $n_{\lambda}=6$ 
the number of wavelength filters and $n_{par}=3$ the number of 
model parameters.

In recent studies it is found that the dominant uncertainty in fitting
fluxes observed with AIA/SDO comes from the incomplete knowledge of the
AIA response functions, which concerns missing atomic lines in the
CHIANTI code as well as uncertainties whether photospheric or coronal 
abundances of chemical elements are more appropriate. The combined
uncertainty is estimated to be $\approx 10-25\%$ of the 
observed AIA fluxes in each wavelength (Boerner et al.~2014; Testa
et al.~2012; Aschwanden et al.~2015).
This is much more than the typical uncertainty due to
photon count statistics, which is of order $\approx 10^{-4}-10^{-3}$ 
for typical AIA count rates ($\approx 10^6-10^8$ DN s$^{-1}$) during flares 
(Boerner et al.~2014; O'Dwyer et al.~2010). We thus
perform the DEM forward-fitting to the 6 wavelength fluxes 
by using the empirical 10\% error of the response functions as an 
estimate of the flux uncertainties due to calibration and background
subtraction errors,
\begin{equation}
	\sigma_{\lambda} \approx 0.1 \ f_{\lambda}^{obs} \ .
\end{equation}

\subsection{	   Spatial Synthesis of Gaussian DEM Fitting 		}

The choice of a suitable DEM function in forward-fitting methods 
is almost an art.
A Gaussian function (in the logarithm of the temperature) appears
to be a good approximation near the peak temperature $T_p$ of 
most DEM functions, but cannot represent ``shoulders'' of the primary 
peak, or secondary peaks at lower or higher temperatures.
This is particularly a problem for
EUV images that have many different temperature structures with
competing emission measures in different areas of the image,
such as multiple cores of hot flaring regions, surrounded by
peripheral cooler regions. 
Therefore it is a sensible approach in the DEM 
parameterization to subdivide the image area of a flare into macropixels
or even single pixels, and then to perform a forward-fit of a
(single-Gaussian) DEM function in each spatial location separately,
while the total DEM distribution function of the entire flare area
can then be constructed by summing all DEM functions from each
spatial location, which we call the ``Spatial Synthesis DEM'' method.
This way, the Gaussian approximation of a DEM
function is applied locally only, but can
adjust different peak emission measures and temperatures at each
spatial location. Such a single-pixel algorithm for automated
temperature and emission measure analysis has been developed
for the 6 coronal AIA wavelength filter images in Aschwanden et al.~(2013),
and a SSW/IDL code is available online 
({\sl http://www.lmsal.com/$\sim$aschwand/software/aia/aia$\_$dem.html}).
The flux $F_{\lambda}(x,y,t)$ is then measured in each pixel 
location $[x,y]$ and time $t$, and the fitted DEM functions 
are defined at each location $[x,y]$ and time $t$ separately,
\begin{equation}
	DEM(T;x,y,t) = EM_p(x,y,t) \exp{ \left(
	- {[log(T) - log(T_p[x,y,t])]^2 \over 2 w_T^2[x,y,t] }
	\right)} \ ,
\end{equation}
and are forward-fitted to the observed fluxes $F_{\lambda}(x,y,t)$ 
at each location $(x,y)$ and time $t$ separately,
\begin{equation}
	F_{\lambda}(x,y,t) - B_{\lambda}(x,y,t) =
	\int DEM(T; x,y,t)  R_{\lambda}(T) dT
	= \sum_{k=1}^{n_T} DEM(T_k; x_i ,y_j,t ) R_{\lambda}(T_k) \Delta T_k \ .
\end{equation}
The synthesized differential emission measure distribution $DEM(T)$
can then be obtained by summing up all local DEM distribution functions
$DEM(T; x,y,t)$ (in units of cm$^{-3}$ K$^{-1}$),  
\begin{equation}
	DEM(T,t) = \int\int DEM(T; x,y,t)\ dx dy 
                 = \sum_{i,j} DEM(T_k; x_i, y_j, t)\ dx_i dy_j \ ,
\end{equation}
and the total emission measure of a flaring region is then obtained
by integration over the temperature range (in units of cm$^{-3}$),
\begin{equation}
	EM(t)   = \int DEM(T,t)\ dT =
		  \sum_k DEM_k(T_k, t) \Delta T_k  \ .
\end{equation}
Note that the synthesized DEM function $DEM(T)$ (Eq.~9) generally
deviates from a Gaussian shape, because it is constructed from the
summation of many Gaussian DEMs from each pixel location with
different emission measure peaks $EM_p(x_i, y_j)$, peak temperatures
$T_p(x_i, y_j)$, and thermal widths $w_T(x_i, y_j)$. This
synthesized DEM function can be arbitrarily complex and accomodate
a different Gaussian DEM function in every spatial location $(x_i, y_j)$.

Typically we process images with a field-of-view of $FOV=0.35$ solar
radius, which corresponds to about 520 AIA pixels. Subdividing these
images into macropixels with a bin size of 4 full-resolution pixels,
we have a grid of 130$\times$130 macropixels $[x_i, y_j]$ and perform
$130^2=16,900$ single-Gaussian DEM fits per time frame, per wavelength set,
and per event.
We illustrate the spatial synthesis procedure with single-Gaussian
DEMs in Fig.~1, where we can see that the local temperature discrimination
yields a higher temperature contrast for increasingly smaller macropixels,
from $N_{bin}=512, 256, 128, 64, 32, 16, 8, 4$ down to 2 image pixels.  
The convergence of the DEM with decreasing bin size is depicted in Fig.~2, 
for three different times of a flare. The initial single-Gaussian DEM 
function fitted to the fluxes of a $512 \times 512$ pixel area (blue curves in
Fig.~2) converges to a double-peaked DEM at the flare peak time (red curve
in middle panel of Fig.2), synthesized from $2 \times 2$ macropixels, 
which evolves then into a broad single-peaked DEM in the postflare
phase (red curve in bottom panel of Fig.~2).

\subsection{		Flare Geometry 				}

The total emission measure EM of a flaring active region, 
such as defined for a single Gaussian DEM (Eq.~2) or for a spatially 
synthesized DEM as defined in Eq.~(10), yield the product of the
squared mean electron density times the flare volume. If we can estimate
the flare volume $V$ from the imaging information, we can then infer
the mean electron density (for unity filling factor). There are many
ways to measure a flare area. Two major problems are the choice of 
a suitable wavelength (in multi-temperature data), and secondly the
choice of a threshold, especially in flares that have a large dynamic
range of fluxes over several orders of magnitude.  

In order to eliminate
the choice of wavelengths, we use the emission measure maps 
$DEM_p(T; x,y,t)$ (Eq.~7), where we find a range of 
$DEM_p = 8.2 \times 10^{22} - 2.7 \times 10^{25}$ cm$^{-5}$ K$^{-1}$
for the peak values. We choose an emission measure threshold near the
lower bound of this range, i.e., $DEM_{p,min}=10^{23}$ cm$^{-5}$ K$^{-1}$,
unless this threshold exceeds the 50\% level of the peak emission measure, 
in which case we use the 50\% level. This way, a flare area is always 
defined, even for flares with low emission measures. We measure then the 
flare area $A$ of thermal emission by counting the number of macropixels 
above the threshold in an emission measure map $EM_p(x,y)$, which 
multiplied with the macropixel size yields an area $A$ (in units of cm$^2$), 
a length scale $L=A^{1/2}$, and a flare volume $V=L^3=A^{3/2}$. 

One problem that we encountered in our analysis is that the flare
area at the peak time $t=t_p$ is sometimes largely inflated due to
saturation of the EUV CCD, pixel bleeding, and diffraction patterns,
and thus no reliable flare area $A_i$ can be measured at the flare 
peak time $t_i$.  Since the automated exposure control alternates 
between short and long exposure times during saturation, an over-exposed 
time frame (with flare area $A_i$ at the peak time $t_i$) 
is interpolated from the preceding time step 
(with flare area $A_ {i-1}$ at time $t_{i-1}$) and the following 
time step (with flare area $A_{i+1}$ at time $t_{i+1}$.
In the derivation of geometric parameters in this study we use the 
maximum flare area $A = max[A(t)]$ measured during the flare duration
interval.

\subsection{		Multi-Thermal Energy 			}

If we substitute the expression of the total emission measure at the
peak time $t_p$ of the flare,
$EM_p = n_p^2 V$, into the expression for the thermal energy $E_{th}$, 
we have the relationship
\begin{equation}
	 E_{th}(t_p) = 3 n_p k_B T_p V   
	             = 3 k_B T_p \sqrt{EM_p \ V} \ .
\end{equation}
This expression is accurate only if the DEM function is a delta-function
with a small thermal width $w_T$, which can then be
characterized by the peak emission measure $EM_p$ at the DEM 
peak temperature $T_p$. 

For every broad temperature DEM distribution $DEM(T)$, as it is 
the case for most solar flares, it is more accurate to perform 
the temperature integral (or summation over discrete temperature
increments $\Delta T_k$, which may be logarithmically binned). 
In the discretized form, the emission measure
$EM_k$ is integrated over the temperature interval $\Delta T_k$ is
$EM_k = DEM(T_k) \Delta T_k$, and the thermal energy can be
written as a summation of partial thermal energies from each
temperature interval $[T_k, T_k + \Delta T_k]$ (see Appendix A),
\begin{equation}
	E_{th} = \sum_k 3 k_B V^{1/2}\ T_k\ EM_k^{1/2}  
	= 3 k_B V^{1/2} \sum_k \ T_k \left[ DEM(T_k)\ \Delta T_k \right]^{1/2} 
		\ .
\end{equation}
While the DEM peak temperatures $T_p$ were
determined within the parameter space of $T_p=0.5-30$ MK, the temperature
integral of the thermal energy (Eq.~12) was calculated in an extended 
range of $log(T_e)=5.0-8.0$, in order to fully include the Gaussian tails
of the DEM fits in each macropixel. This yields a more accurate value
of the total multi-thermal energy, since it avoids a truncation at the
high-temperature tail of the composite DEM distribution.
Note that we defined the thermal energy in terms of the volume-integrated
total emission measure ($EM = \int DEM(T) dT = \int n_e^2 dV = n_e^2 V$
(in units of cm$^{-3}$), in contrast to the column depth integrated emission 
measure per area, $EM/A = \int DEM(T)/A\ dT = \int n_e^2 dz = n_e^2 L$
(in units of cm$^{-5}$) used in the spatial synthesis method (Eq.~7), 
where the emission measure is quantified per unit area or per image pixel
(see detailed derivation in Appendix A).
We find that the more accurate expression of Eq.~(12) typically yields 
a factor of $\approx 14$ higher values for the thermal energies than the 
single-temperature approximation of Eq.~(11), and thus represents a very 
important correction for broad multi-temperature DEMs. 

Considering the more complex DEM functions obtained from spatial synthesis
with Eq.~(9), we will see that the DEM function often has multiple peaks,
and thus it does not make any sense anymore to talk about a single peak
emission measure $EM_p$ and single peak temperature $T_p$. In order to
characterize such complex DEM functions with a characteristic temperature
value, it makes more sense to define an emission measure-weighted
temperature $T_w$, which we define as,
\begin{equation}
	T_w = { \int T \ DEM(T) dT \over \int DEM(T) dT }
	= { \sum_k T_k \ DEM(T_k) \Delta T_k \over EM } \ .
\end{equation}
and approximately characterizes the ``centroid'' of the DEM function.

\section{ 		OBSERVATIONS AND RESULTS 		}

\subsection{		AIA Observations			}

The dataset we are analyzing for this project on the global energetics
of flares includes all M- and X-class flares observed with SDO
during the first 3.5 years of the mission (2010 June 1 to 2014 Jan 31),
which amounts to 399 flare events, as described in Paper I (Aschwanden, Xu, 
and Jing 2014a). The catalog of these flare events is available online, see
{\sl http://www.lmsal.com/$\sim$aschwand/RHESSI/flare$\_$energetics.html}.
We attempt to calculate the thermal energies in all 399 catalogued events,
but we encountered 8 events with incomplete or corrupted AIA data,
so that we are left with 391 events suitable for thermal data analysis.

AIA provides EUV images from four $4096 \times 4096$ detectors
with a pixel size of $0.6\arcsec$, corresponding to an effective
spatial resolution of $\approx 1.6\arcsec$. We generally use a
subimage with a field-of-view of $FOV=0.35\ R_{\odot}$. AIA records a full set of
near-simultaneous images in each temperature filter with a fixed cadence
of 12 seconds, while our analysis of the flare evolution is done in
time increments of $\Delta t=0.1$ hrs. This cadence may underestimate
the maximum thermal energy during a flare in some cases, but is estimated 
to be less than a factor of 2.  

\subsection{		Example of DEM Analysis		}

An example of our DEM analysis is summarized in Fig.~3,
which applies to the first event (\# 1) of our list, a GOES M2.0
class flare in active region NOAA 11081 at N23 W47, observed with AIA/SDO 
on 2010 June 12, 00:00-01:30 UT. The GOES 1-8 \ang\ light curve is
shown in Fig.3a, with GOES flare start time at $t_s$=00:30 UT, peak time
at $t_p$=00:58 UT, and flare end time at $t_e$=01:02 UT, according to 
the NOAA event list. The flare end time $t_e$ is defined when the GOES flux 
drops down to 50\% of the peak value, according to NOAA convention,
but flare-related EUV emission always lasts significantly longer.
In our thermal analysis we add margins of $\Delta t=0.5$ hr before
and after the NOAA flare start and end times, which covers the
time interval of 00:00-01:32 UT in this event. We use a cadence
of $dt=0.1$ hr, which yields 14 time frames for this event.
The 6 AIA flux profiles are shown in Fig.~3b, which show a very
simple evolution of a single peak in all 6 EUV wavelengths, coincident
with the SXR peak in GOES time profiles. 
The peak time occurs in time frame $i_t=10$, at 
00:58 UT. Flare background fluxes have been subtracted in
every spatial macropixel ($4 \times 4$ image pixels) separately
(according to Eq.~8). Because every macropixel has a different
background value $B_{\lambda}(x,y,t_b)$, the summation of all
background subtracted profiles $F_{\lambda}(x,y,t)-B_{\lambda}(x,y,t)$
leaves residuals that amount to a fraction of $\approx 0.05-0.5$ of
the peak flux (see preflare time profile of spatially-summed fluxes
in Fig.~3b).

For the DEM analysis we read 14 (time frames) times 6 (wavelength)
AIA images, extract subimages within a $FOV=0.35$ solar radii, which
amount to a size of about 522 pixels, we rebin the images into 
$4\times 4$ macropixels,
yielding a spatial 2D array $(x_i, y_j)$ of $130 \times 130$ macropixels, 
subtract in each macropixel a temporal minimum flux background, 
forward-fit a Gaussian DEM function in each macropixel, which yields the 3 
Gaussian parameters: the DEM peak emission measure $EM_p(x_i, y_j)$,
DEM peak temperature $T_p(x_i, y_j)$, and thermal width 
$w_T(x_i, y_j)$, or a Gaussian DEM function $DEM(T; x_i, y_j)$
(Eq.~7) for each macropixel. Summing the $130 \times 130 = 16,900$
single-Gaussian DEMs yields then a spatially synthesized DEM function
that is shown in Fig.~3f for each time step $i_t=1,...,14$.
The evolution of the DEM peak starts from a DEM peak
temperature of $T_p(i_t=1)=10^{6.4}=2.5$ MK and peaks at a value of
$t_p(i_t=10)=10^{6.8}=6.3$ MK, and decreases again to the preflare
value. The evolution of this peak temperature $T_p(t)$ is also shown
in Fig.~3c, along with the evolution of the 
mean temperature $T_e(t)$, the mean electron density 
$n_e(t)=\sqrt{EM_p(t)/V}$, and the thermal energy $E_{th}(t)$ (Eq.~12),
in normalized units. The spatial distribution of the emission measure
map $EM_p(x_i, y_j)$ is shown in Fig.~3e, where instrumental
diffraction patterns (diagonal features) and pixel bleeding (vertical
feature) are visible also at the flare peak time. 
Since these instrumental effects are mostly
a spatial re-distribution of photons inside the FOV of the observed
image, we expect that they do not affect much the obtained DEM 
function after spatial integration. The emission measure maps serve
to measure a wavelength-independent flare area $A$ at the flare peak
time (above some threshold; Section 2.4), which yields the equivalent
length scale $L=A^{1/2}$. The physical parameters obtained
for this event at the flare peak time are listed in Fig.~3 (bottom
right). Note that the peak temperature is only $T_p=6.31$ MK, while
the emission measure-weighted temperature $T_w=18.57$ MK (Eq.~13)
is substantially
higher. The flare length scale (indicated with a square in Fig.~3e)
is $L=13.2$ Mm, the electron density is $n_e=\sqrt{EM_p/V}=5.8 \times
10^{10}$ cm$^{-3}$, and the thermal energy is $E_{th}=7.0 \times 10^{30}$
erg for this event.

The goodness-of-fit or reduced $\chi^2$-criterion of the DEM fit yields 
a mean and standard deviation of $\chi^2=0.24 \pm 0.43$ for the 14 DEM fits 
of this particular event \#1 (Fig.~3). As mentioned before (Section 2.2),
the calculation of the reduced $\chi^2$-criterion is based on the estimated
uncertainty of the observed AIA fluxes, which is dominated by the incomplete 
knowledge of the instrumental response functions, estimated to be of order
$\approx 10-25\%$ (Boerner et al.~2014; Testa et al.~2012). Although 
the $\chi^2$-value found for this particular event is relatively low,
compared with the mean statistical expectation, it fits into the broad
range of the obtained overall statistical distribution. In Fig.~6g we
plot the distribution of the $\chi^2$-values of the 391 fitted
flare events, where the $\chi^2$ value of each flare event is a
time average, as well as a spatial average (using the spatial synthesis
method). The peak of this distribution is near $\chi^2 \approx 1$ and
the median is $\chi^2 \approx 1.3$ (Fig.~6g), which indicates that the   
chosen model of the DEM parameterization (Eq.~1) yields a best fit that is
consistent with the empirical estimates of uncertainties in the flux or 
response functions (Eq.~6). Of course, the Gaussian DEM parameterization,
even when individually fitted in each pixel, may not always represent
the best functional form of observed DEMs, which may explain some
$\chi^2$-values significantly larger than unity.
A more accurate goodness-of-fit test would 
require a more complex parameterization of the DEM function and a
physical model of the flux uncertainties $\sigma_\lambda$, 
which should include systematic uncertainties due to the AIA flux 
calibration, the atomic (coronal and photospheric) abundances, the 
atomic transitions (computed with the CHIANTI code here), and the 
background subtraction method, which is not attempted here.

\subsection{	DEM Functions of Extreme Events			}

In Fig.~4 we show the differential emission measure distributions
$DEM(T)$ of 12 extreme events among the 391 analyzed M and X-class
flare events. These
12 events were selected by the minimum and maximum values in the
parameters of the length scale $L$ (Fig.~4a,b), the DEM peak
temperature $T_p$ (Fig.~4c,d), the emission measure-weighted
temperature $T_w$ (Fig.~4e,f), the electron density $n_e$ 
(Fig.~4g,h), the DEM peak emission measure $EM$( Fig.~4i,j),
and flare duration $D$ (Fig.~4k,l). 
This selection of extreme events demonstrates the
variety and diversity of DEM functions we encountered among the
analyzed flare events. It shows also the versatility and adequacy of the 
DEM parameterization using spatially synthesized (single-Gaussian) 
DEM functions. 

The length scales of thermal emission vary from $L_{min}=1.7$ Mm
(\#256; Fig.~4a) to $L_{max}=45.9$ Mm (\#132; Fig.4b). What is striking
between the evolution of these two events is that the flare with the
smallest size shows very little increase in the emission measure at
any temperature, while the largest flare exhibits a large 
increase in the high-temperature emission measure. 

For the peak temperatures we find a range from $T_p=0.5$ MK
(\#305; Fig.~4c) to $T_p=28.1$ MK (\#67; Fig.~4d), which is
not necessarily coincident with the emission measure-weighted
temperature $T_w$. This is clearly shown in the case with
the smallest peak temperature,
which is far below the emission measure-weighted temperature
of 
peaks in the DEM, which can make the peak temperature to jump
around wildly as a function of time, as long as their associated DEM 
peak emission measures are comparable. This is a major reason why the
DEM peak temperature should not be used in the estimate
of thermal energies, but rather the emission measure-weighted
temperature that is a more stable characteristic of the DEM
function.

For the emission measure-weighted DEM function we find a 
range from $T_w=5.7$ MK for the coldest flare (\#102; Fig.~4e) 
to $T_w=41.6$ MK for the hottest flare (\#316; Fig.~4f), 
which is close to the upper limit of
the temperature range where AIA is sensitive.
The coldest flare in our selection
with $T_w=5.7$ MK is a M1.3 GOES class, 
while the hottest flare with $T_w=41.6$ MK is a M3.5 
GOES class. The GOES class does not necessarily
correlate with the flare temperature, which is expected
since the GOES class is mostly defined by the emission 
measures (in soft X-rays) rather than by the temperature.

For the electron density we find a range from
$n_e=10^{10.31}$ cm$^{-3}$ (\#396; Fig.~4g) to
$n_e=10^{11.77}$ cm$^{-3}$ (\#375; Fig.~4h), which
corresponds to a variation by a factor of $\approx 30$. 
The lowest density corresponds to a low peak 
temperature ($T_p=1.6$ MK), while the 
highest density yields a high peak temperature
temperature ($T_p=29.5$ MK).
For a fixed loop length, a correlation between the electron density
and the electron temperature is expected according to the RTV scaling law,
i.e., $n_p \propto T_p^2$ (Eq.~21).

For the DEM peak emission measure we find a variation from
$EM_p=10^{47.31}$ cm$^{-5}$ (\# 241; Fig.~4i) to
$EM_p=10^{50.26}$ cm$^{-5}$ (\# 147; Fig.~4j), which
varies by a factor of $\approx 1000$. The corresponding
GOES classes are M1.3 and X5.4, which are both near the
limits of the GOES class range (M1.0 - X6.9) found in our selection.
The event with the largest emission measure represents 
the second-largest GOES class (X5.4) in our selection, 
and thus the GEOS class is indeed a good proxy to 
estimate the emission measure of flares.  

The time range of flare durations is found to vary from
$D=0.1$ hr (\#56; Fig.~4k) to $D=4.1$ hr (\#130; Fig.~4l).
The longest duration event, however, does not have extreme
values in temperature, emission measure, or length scale.

\subsection{	Statistics of Physical Parameters 		}

We provide some statistics on the derived thermal parameters,
such as the length scale $L$, the thermal volume $V$,
the DEM peak temperature $T_p$, the emission measure-weighted 
temperature $T_w$, the electron density $n_e$, the total
emission measure $EM$, and the thermal energy $E_{th}$, 
in form of scatterplots (Fig.~5) and size distributions (Fig.~6).
The inferred physical parameters are listed for the 28 X-class
flares in Table 1, and for all 391 M and X-class flares in the
machine-readable Table 2.
The ranges of these physical parameters have already been
discussed in terms of extreme values in Section 3.3. 
The scatterplots shown in Fig.~5
reveal us which parameters are correlated and indicate
simplified scaling relationships, while the size distributions
shown in Fig.~6 reveal us the powerlaw tails that are typical for
dissipative nonlinear systems governed by self-organized criticality (SOC).

The scatterplots shown in Fig.~5 indicate that the thermal energy
is correlated with the length scale $L$ by the scaling relationship
(Fig.~5a),
\begin{equation}
	E_{th} \propto L^{2.3 \pm 0.1} \ ,
\end{equation} 
and consequently is correlated with the volume $V$ (Fig.~5b) also,
\begin{equation}
	E_{th} \propto V^{0.76 \pm 0.04} \ ,
\end{equation} 
and is correlated also with the total emission measure $EM$ (Fig.~5f)
\begin{equation}
	E_{th} \propto EM^{1.27 \pm 0.10} \ ,
\end{equation} 
but strongly anti-correlated with the electron density $n_e$ (Fig.~5e),
and is not correlated with the temperatures $T_p$ (Fig.~5c)
and $T_w$ (Fig.~5d).

Regarding the size distributions, the {\sl fractal-diffusive self-organized
criticality (FD-SOC)} model provides predictions for the size distributions
(Aschwanden 2012; Aschwanden et al.~2014b). 
The most fundamental parameter in the FD-SOC model is
the length scale $L$, which according to the scale-free probability
conjecture is expected to have a size distribution $N(L) \propto L^{-d}$
for Euclidean space dimension $d$. We find agreement between
this theory and the data within the uncertainties of the fit (Fig.~6a),
\begin{equation}
	\alpha_L^{obs} = 3.3 \pm 0.3 \ , \qquad 
	\alpha_L^{theo} = 3.0 \ . 
\end{equation}
For the volume $V$ of thermal emission, the FD-SOC model predicts
a powerlaw slope of $\alpha_V=1+(d-1)/d$, and we find good agreement (Fig.~6b),
\begin{equation}
	\alpha_V^{obs} = 1.7 \pm 0.2 \ , \qquad 
	\alpha_V^{theo} = 1.67 \ . 
\end{equation}
For the energy $E$, using the observed scaling, i.e., 
$E_{th} \propto V^\gamma$ with $\gamma=0.76$ (Eq.~15), we expect then 
a size distribution of $N(E_{th}) dE_{th} \propto N[V(E_{th})] 
|dV/dE_{th}| dE_{th} \propto E_{th}^{-(1 + (2/3)\gamma)}$, which
predicts a powerlaw slope of $\alpha_{Eth}=[1+(2/3)\gamma] \approx 1.88$,
which is indeed consistent with the observed slope,
\begin{equation}
	\alpha_E^{obs} = 1.8 \pm 0.2 \ , \qquad 
	\alpha_E^{theo} = 1.88 \ . 
\end{equation}
We have to keep in mind that the FD-SOC model is a very generic
statistical model that predicts a universal scaling law for spatial
parameters, based on the scale-free probability conjecture, i.e.,
$N(L) \propto L^{-D}$ (Aschwanden 2012), while the scaling of
other physical parameters, such as the energy, $E_{th} \propto V^\gamma$,
requires a physical model that is specific to each SOC phenomenon.
In the next Section we will discuss the RTV scaling law, which we
apply to model the otherwise unknown scaling of the energy with
the volume, $E_{th} \propto V^\gamma$.

\subsection{	The Rosner-Tucker-Vaiana Scaling Law 		}

A well-known physical scaling law between hydrodynamic parameters
of a coronal loop is the Rosner-Tucker-Vaiana law (Rosner et al.~1978),
which is derived under the assumption of energy balance between the
energy input by a volumetric heating rate $E_h$ (in units of
[erg cm$^{-2}$ s${-1}$]) and the radiative $E_R$ and the conductive 
loss rates $E_C$, i.e., $E_H - E_R - E_C = 0$, which yields two
scaling laws between the loop length $L$, loop apex electron temperature $T_e$,
average electron density $n_e$, and heating rate $E_H$. While this 
original derivation applies to a steady-state of a heated coronal loop,
it turned out that the same scaling laws apply also to solar flares
at the heating/cooling turnover point (Aschwanden and Tsiklauri 2009). 
Solar flares
are generally not heated under steady-state conditions, except
at the turning point of maximum temperature, when the heating rate and
the radiative and conductive losses are balanced for a short instant
of time. Before reaching this turning point, heating dominates the 
cooling losses, while the cooling dominates after this turning point. 

We can express the RTV scaling laws explicitly for the parameters
$T_e, n_e, L, EM, E_{th}$ (Aschwanden and Shimizu 2013),
\begin{equation}
        T_{RTV} = c_1 \ n_e^{1/2} \ L^{1/2} \ , \qquad
        c_1 = 1.1 \times 10^{-3} \ ,
\end{equation}
\begin{equation}
        n_{RTV} = c_2 \ T_e^{2} L^{-1} \ , \qquad
        c_2 = 8.4 \times 10^5 ,
\end{equation}
\begin{equation}
        L_{RTV} = c_3 \ T_e^{2} n_e^{-1} \ , \qquad
        c_3 = 8.4 \times 10^5 \ .
\end{equation}
\begin{equation}
        EM_{RTV} = \int n_e^2 dV = n_e^2 V = n_e^2 ({2 \pi \over 3} L^3 )
             = c_4 \ T_e^4 L \ , \qquad
               c_4 = 1.48 \times 10^{12} .
\end{equation}
\begin{equation}
        E_{th,RTV} = 3 n_e k_B T_e V
                = c_5 \ T_e^3 L^2 \ , \qquad
                  c_5 = 7.3 \times 10^{-10} .
\end{equation}
We can then compare the observed parameters 
$T_e, n_e, L, EM, E_{th}$ with these theoretically predicted
parameters $T_{RTV}, n_{RTV}, L_{RTV}, EM_{RTV}, E_{th,RTV}$,
which is shown in Fig.~7. Note that we use the weighted temperature
$T_w$ and the emission measure $EM_p$ and density $n_p$ measured
at the peak time $t_p$ of the flare here.
While the original RTV scaling law has no
free parameters, the scaling between the average loop half length $L_{loop}$
(required for the RTV scaling law) and the average length scale $L$ 
(measured here during the flare duration) requires a geometric model,
as well as information on filling factors and fractal geometry.
Since detailed modeling of the 3D geometry of flare loop configurations
is beyond the scope of this study, we determine the average scaling ratio 
empirically and find that a relationship of $L \approx (2 \pi) \ L_{loop}$ 
yields a satisfactory match between the observed and the
theoretically predicted physical parameters of the RTV scaling law
(indicated with the dotted diagonal line expected for equivalence
in Fig.~7).  

We see now that the 3-parameter RTV scaling laws (Fig.~7) retrieve 
the relationships obtained from 2-parameter correlations (Fig.~5).
The correlation of the thermal energy with length scale,
$E_{th} \propto L^{2.3\pm0.1}$ (Eq.14; Fig.~5a) is similar
to the RTV relationship $E_{th} \propto L^2$ (Eq.~24),
which is equivalent to the relationship with the volume, i.e.,
$E_{th} \propto V^{0.76\pm0.04}$ (Eq.15; Fig.~5b) and 
the RTV relationship $E_{th} \propto L^2 \propto V^{2/3}$ (Eq.~24).
Combining the RTV relationships between $E_{th}$ (Eq.~24) and
$EM$ (Eq.~23) we obtain $E_{th} \propto EM_p (L/T)$, which
is {\sl similar} to the observed 2-parameter correlation 
$E_{th} \propto EM^{1.3 \pm 0.1}$. Thus the 2-parameter
correlations are approximations of the 3-parameter (RTV) scaling laws,
and thus can be explained by a physical model, although they 
are less accurate because of the neglected third parameter.
Comparing the observed and RTV-predicted values (as shown in Fig.~7),
we find that the (multi-)thermal energies $E_{th,RTV}$, emission
measures $EM_{RTV}$, and length scales $L_{RTV}$  are correlated
with the observed values within a standard deviation, while the
temperature $T_{RTV}$ and density $n_{RTV}$ deviate more than a
standard deviation, which is likely to be caused by their smaller
ranges of values and the associated truncation effects (e.g.,
see calculation of truncation effects in Fig.~8 of Aschwanden
and Shimizu 2013).

\subsection{	Comparison of Magnetic and Thermal Energies	}

The main goal of the global flare energetics project is the
comparison and partitioning of various flare energies. In Paper I we
calculated the dissipated magnetic energies in 172 M and X-class
flares, based on the (cumulative) decrease of free energies during
each flare, which were found to have a range of 
$E_{diss}= ( 1.5 - 1500 ) \times 10^{30}$ erg. In this study we
calculated the thermal energy at the peak time of the total
emission measure and find a range of 
$E_{th}= ( 0.15 - 315 ) \times 10^{30}$ erg. A scatterplot 
between the magnetic
and thermal energies is shown in Fig.~8a. From this diagram we
see that the average ratio is $E_{th}/E_{diss} \approx 0.082$,
with a standard deviation by a factor of 4.8, which defines a typical
range of $E_{th}/E_{diss}=0.02-0.40$. Thus, the thermal energy
amounts generally only to a fraction of $\approx 2\%-40\%$ of the
dissipated magnetic energy, as determined with the coronal NLFFF
method.

We show also a scatterplot of the thermal energy with the
dissipated magnetic energy as computed with the photospheric NLFFF 
method, which could be performed only for 12 events (Fig.~8b).
In this small dataset, the average ratio is $q_e=0.76$, with a scatter
by a factor of 6.5, or a range of $q_e \approx 0.12-4.8$. In four out
of the 12 events the thermal energy exceeds the dissipated magnetic
energy, which is likely to be a false result due to underestimates
of the dissipated magnetic energy, since the PHOT-NLFFF code seems
to be less sensitive in measuring decreases of the free energy
than the COR-NLFFF code, possibly due to a smoothing effect caused
by the preprocessing procedure. 

We compare the new results also with the previous study by Emslie
et al.~(2012), where the thermal energy could be determined for
32 large eruptive flares, while the magnetically dissipated energy
was estimated to be 30\% of the potential energy. In that study,
the average ratio of the thermal to the magnetically dissipated
energy is found to be $E_{th}/E_{diss} \approx 0.0045$ with a
scatter by a factor of $\approx 2.3$, which yields a range of
0.2\%-1.0\% (Fig.~8c). Since the thermal energies have a similar
median value ($E_{th,med}=4.6 \times 10^{30}$ erg) as we find in this study
($E_{th,med}=6.0 \times 10^{30}$ erg), the discrepancy is most likely
attributed to an overestimate of the magnetically dissipated 
energies, as well as to a selection effect of larger flares. 
The median value of the magnetically dissipated energy is $E_{diss,med}=1300 
\times 10^{30}$ erg in Emslie et al.~(2012), while we find a
median value of $E_{diss,med}=110 \times 10^{30}$ erg, which
is about an order of magnitude lower, and goes along with our
finding that the free energy is about 1\%-25\% of the potential
free energy, rather than 30\% as assumed in the study of Emslie
et al.~(2012).

\section{ 		DISCUSSION 				  }

\subsection{	Previous Measurements of Thermal Flare Energies	  }

Most previous studies estimated thermal flare energies by using the 
isothermal relationship, i.e., $E_{th} = 3 k_B T_p \sqrt{EM_p \ V}$,
which requires a DEM analysis (to obtain the peak emission measure
$EM_p$ and peak temperature $T_p$) and imaging observations
(in order to obtain the flare area or volume $V$), measured at the
flare peak time. A DEM analysis requires multiple temperature filters,
and thus thermal flare energies can only be obtained from instruments
with multi-wavelength imaging capabilities. Statistics of thermal 
energies was gathered for large flares, nanoflares, and impulsive
brightenings in EUV and soft X-rays from 
{\sl Skylab S-054} (Pallavicini et al.~1977), {\sl Yohkoh/SXT} 
(Shimizu 1997; Aschwanden and Benz 1997; Shimojo and Shibata 2000), 
SoHO/EIT (Krucker and Benz 2000); TRACE (Aschwanden et al.~2000; 
Aschwanden and Parnell 2002), RHESSI (Emslie et al.~2004, 2005, 
2012; Caspi et al.~2014), and AIA/SDO (Aschwanden and Shimizu 2013). 

How consistent are the thermal energies determined here with previous 
measurements ? We compile some statistics on thermal energy measurements 
in large flares in Table 3, by listing the instruments, the number of 
events, and the parameter ranges of the spatial 
scale $L$, the peak electron temperature $T_p$, the peak electron density 
$n_p$, the peak emission measure $EM_p$, and the thermal energy $E_{th}$.  
A scatterplot of thermal energies $E_{th}(V)$ versus the flare volumes $V$ 
measured in large flares is shown in Fig.~9.
In particular, statistics on large flares (approximately
GOES M- and X-class) has been analyzed in 31 events from Skylab S-054
(Pallavicini et al.~1977), in 32 events from RHESSI (Emslie et al.~2012),
in 155 events from AIA/SDO (Aschwanden and Shimizu 2013), and in 391 events
from AIA/SDO in the present study. Table 3 provides the ranges of
reported physical parameters, but we have to be aware that different
event selections have been used in the different datasets.

\subsection{	Isothermal Versus Multi-Thermal Energies	}

The most striking discrepancy appears between the isothermal and
multi-thermal energies, which is measured for the first time in this study.
We overlay the thermal energies $E_{th}$ as a function of the 
flare volume $V$ for the same four studies in Fig.~9. In the present study 
we calculate both the isothermal energy $E_{th,iso}$ (Eq.~11) and the 
multi-thermal energy $E_{th,multi}$ (Eq.~12) and find a systematic
difference of $E_{th,multi}/E_{th,iso} \approx 14$ (Fig.~9, 10). 
Note the offset of the linear regression fits between isothermal energies 
(black line and diamonds in Fig.~9) and multi-thermal energies (orange line 
and diamonds in Fig.~9).
The multi-thermal flare energy definition has to our knowledge not 
been applied in the calculation of thermal flare energies in all
previous studies, but is very important, because it boosts the thermal
energy produced in flares statistically by an average factor of $\approx 14$,
as measured from the energy offset in cumulative size distributions 
(Fig.~10). This is related to the incompatibility of iso-thermal temperatures 
inferred from GOES, AIA, and RHESSI data, investigated in a
recent study (Ryan et al.~2014), which can only be ameliorated with 
broadband (multi-temperature) DEM distributions.
The systematic underestimate of the thermal energy, when the isothermal
approximation is used, may also be the reason why a very low value of
$E_{th}/E_{diss}=0.2\%-1\%$ (Fig.~8c) was found for the thermal/magnetic
energy ratio in Emslie et al.~(2012), compared with our range of 
$E_{th}/E_{diss}=2\%-40\%$ (Fig.~8a) calculated in the present study. 

\subsection{	Flare Volume Measurements			}

The thermal energy depends on the volume $V$, and thus the measurement
of flare areas or volumes are crucial to obtain an accurate energy value.
Since we can directly observe in 2D images the flare area $A$ only, the
definition of a flare volume $V$ is subject to modeling. The simplest
definition is the Euclidean relationship $V=A^{3/2}$ and $L=A^{1/2}$,
but more complicated definitions involve the fractal dimension
(Aschwanden and Aschwanden 2008a,b), 3D filling factors (Aschwanden
and Aschwanden 2008b), or other geometric concepts to characterize the 
inhomogeneity of flare plasmas. One prominent modeling concept is the
hydrostatic density scale height $\lambda(T)$, which depends on the
flare plasma temperature $T$ 
and can be used to estimate the vertical height above the
solar surface. The detailed geometry of the flare plasma often appears
to have the geometry of an arcade of loops, which can be highly 
inhomogeneous, depending on the spatial intermittency of precipitating
electrons along the flare ribbons. Nevertheless, regardless how
complicated the spatial topology of a flare is, the thermal energy
is a volume integral and thus should be rotation-invariant to the 
aspect angle or heliographic location (assuming that we measure
correct DEMs along each line-of-sight). This argument justifies 
isotropic geometries such as hemispheric flare volumes 
(Aschwanden and Shimizu 2013), or the related Euclidean relationship 
$V \approx L^3$. Moreover, the height $h=L/2$ of semi-circular flare 
loops is about half of the footpoint separation $L$, and thus the 
volume $V=L^2 h = (L^3)/2$ can be approximated with a cube $V \approx L^3$.
Hence, we use the simple Euclidean relationships $V=A^{3/2}$ and 
$L=A^{1/2}$ in this paper. Detailed geometric 3D modeling of the flare 
volume at different temperatures is beyond the scope of this study.

How consistent is the flare volume measurement in the present work 
with previous studies? Most flare area measurements are done using a
flux threshold, which is chosen above the data noise level and lower
than the maximum flux in an image, but is arbitrary within this range.
The volume of limb flares from Skylab data (Pallavicini et al.~1977) 
was calculated from measuring the height and size of bright soft X-ray
emission in photographs and yields a remarkable good match for the
isothermal energy with our present study (blue line and crosses
in Fig.~9).  The previous study of 155 M- and X-class flares with 
AIA/SDO data (Aschwanden and Shimizu 2013) involved multiple flux
threshold levels and was combined from 6 different wavelength filters,
but is consistent with the area measurements in this study within
a factor of $\lapprox 2$. This uncertainty translates into a factor
of $2^{3/2} \approx 3$ for volumes, total emission measures, and
thermal energies.

\subsection{	Spatial-Synthesized DEM Analysis 		}

DEM analysis is a prerequisite tool for determining thermal energies.
The thermal width of the DEM distribution is the most crucial 
parameter to discriminate between isothermal and multi-thermal cases.
Numerical integrations of the DEM temperature distribution show that
multi-thermal DEMs yield in the average 14 times higher (multi-)thermal
energies than isothermal (delta-function-like) DEM distributions
(Fig.~10). Thus, the fidelity
of the DEM reconstruction is important for the accurate determination 
of thermal flare energies. 

In this study we employed the spatial synthesis DEM method 
(Section 2.3 and Aschwanden et al.~2013),
which approximates the DEM in every (macro-)pixel with a 3-parameter
Gaussian DEM function, which is then synthesized for the entire flare
volume by adding all partial DEM distributions from each pixel. 
In Fig.~2 we demonstrated that this method
converges to a unique DEM solution by iterating from large macro-pixels
to smaller sizes, down to a single image pixel. We find that this method
converges rapidly, when iterating macro-pixel sizes $\Delta x = X \times 2^{-i}$, 
$i=0,...,8$, on an image with full size $X$ (Figs.~1 and 2). This means 
that macropixels with a size of a few pixels
isolate hot flare areas and ambient cooler plasma areas sufficiently 
to be characterized with a single-peaked DEM function. The fast convergence
to a unique DEM function is very fortunate and relieves us from more 
sophisticated DEM modeling. 

We find that the largest uncertainty in DEM modeling comes
from uncertainties of the instrumental response functions,
including missing atomic lines, chemical abundance variations,
and preflare-background subtraction, which all combined are 
estimated to be of order $\approx 10-25\%$ (Boerner et al.~2014; 
Testa et al.~2012; Aschwanden et al.~2015), which is also confirmed
from DEM inversions applied to synthetic data generated with 
3D magneto-hydrodynamic (MHD) simulations (Testa et al.~2012). 

\subsection{	Scaling Law and Extreme Events 			}

In Section 3.5 we derived a physical scaling law for the thermal
energy, $E_{th,RTV} = 7.3 \times 10^{-10} \ T_p^3 L_p^2$ (Eq.~24), 
based on the RTV scaling law of 1-D hydrostatic loops that are in 
steady-state energy balance between heating and cooling processes. 
The observational measurements of (multi-thermal) energies were found
indeed to match this predicted relationship closely (see correlation
between theoretically predicted and observed thermal energies in
Fig.~7e). 

Let us consider the parameters of the most extreme events.
For the largest flare in our dataset, we found a length
scale of  $L_p = 10^{9.7} \approx 50$ Mm $\approx 0.07$ solar radius, 
the hottest flare has an (emission measure-weighted) temperature of 
$T_p=10^{7.6} \approx 40$ MK, and the most energetic flare has
a multi-thermal energy of $E_{th}= 10^{32.0}$ erg. 
The upper limit for thermal energies is of particular
interest for predictions of the most extreme (and worst events
for space weather and astronauts). Based on the largest flare events
observed in history, with a GOES-class of X10 to X17, an even larger
maximum flare energy of $E_{max} \approx 10^{33}$ erg was estimated,
while stellar flares may range up to $E_{max} \approx 10^{36}$ erg
(see Fig.~3 in Schrijver et al.~2012).

On the other extreme, the RTV scaling law (Eq.~24) may also be applied
to predict the magnitude of the smallest coronal flare events. An
absolute lower limit of flare temperatures is the temperature of
the ambient solar corona, which is approximately $T_{min} \approx
1.0$ MK. For a lower limit of the spatial size of a flare event we
can use the size of the smallest loop that sticks out of the chromosphere,
which has a height of $h_{chrom} \approx 2$ Mm and a semi-circular
loop length of $L_{min}=\pi h_{chrom} \approx 6$ Mm. The apex
segment that sticks out of the chromosphere can have a projected
length scale as short as $L_{min} \gapprox 1$ Mm. The extrapolated
thermal energy of the smallest flare is then estimated to be 
$E_{th}= 7.3 \times 10^{-10} \ T_{min}^3 L_{\min}^2 \approx 
7 \times 10^{24}$ erg, which is about 9 orders of magnitude smaller
than the largest flare, and thus called a nanoflare. This is
consistent with the smallest observed nanoflares, which have been
found to have a thermal energy of $E_{th} \approx 10^{24}-10^{26}$ erg
(Krucker and Benz 2000; Parnell and Jupp 2000; Aschwanden et al.~2000; 
Aschwanden and Parnell 2002). Note that these predictions are based
on our calculations of the multi-thermal energy, which amounts to
an average correction factor of $\lapprox 14$. 

\subsection{	Self-Organized Criticality Models		}

The statistics of nonlinear dissipative events often follows a
scale-free powerlaw distribution, in contrast to (linear) random 
processes (such as photon statistics of a steady source), which
follow a Poisson distribution (or its exponential approximation). 
The powerlaw function
in occurrence frequency distributions (or size distributions)
has been declared as a hallmark of nonlinear systems governed by
self-organized criticality (SOC; Bak et al.~1987). A quantitative
derivation of the powerlaw distribution function of SOC processes
has been derived in the framework of the fractal-diffusive self-organized
criticality model (FD-SOC: Aschwanden 2012, Aschwanden et al.~2014b),
which predicts universal values for the powerlaw slopes of spatio-temporal
parameters, based on the {\sl scale-free probability conjecture},
$N(L) \propto L^{-d}$, the fractal geometry of nonlinear dissipative
avalanches, and diffusive transport of the avalanche evolution.
We measured the size distributions of spatio-temporal physical parameters 
in solar flares (length $L$, area $A$, volume $V$, durations $D$) 
and found indeed agreement with the predictions of the 
standard FD-SOC model (Fig.~6). The size distributions of the other
physical parameters ($T_p$, $n_p$, $EM_p$, $E_{th}$), however, are
not universal, but depend on the underlying physical process of the
SOC phenomenon. For solar flares in particular, we found that the
RTV scaling law is consistent with the observed parameter correlations
and size distributions. Most of the physical scaling laws are expressed
in terms of powerlaw exponents (such as the thermal energy, i.e.,
$E_{th} \propto T_p^3 L_p^2$), which has the consequence that all
size distributions of physical parameters are also predicted to have
a powerlaw shape, except for finite-size effects (that produce a steep
drop-off at the upper end) and incomplete sampling due to limited
sensitivity (which produces a turnover at the lower end), as manifested
in the size distributions shown in Fig.~6. What the observed size
distributions show, is the scale-free parameter range (also called
inertial range) of SOC processes over which an identical physical 
process governs nonlinear energy dissipation. The size distributions 
shown in Fig.~6 exhibit no indication of multiple or broken powerlaws 
in the inertial range of M- and X-class flares. Note that such powerlaw
distributions occur only for statistically complete samples (above
some threshold value). Datasets with ``hand-selected'' events 
(such as the 37 eruptive flare events sampled in Emslie et al.~2012)
do not exhibit powerlaw-like size distributions.

Various flare energy size distributions have been compared in previous
studies (e.g., see composite size distribution in Fig.~10 of Aschwanden
et al.~2000, based on size distributions published by 
Shimizu 1997; Crosby et al.~1993; Krucker and Benz 2000; 
Parnell and Jupp 2000; and Aschwanden et al.~2000).
Such composite size distributions have been used to characterize
the overall size distributions from the smallest nanoflare to the
largest X-class flare. However, the construction of a synthesized
flare energy size distribution requires a consistent definition
of energy, which is not the case in most of the published
studies, since they contain thermal as well as nonthermal energies.
In order to illustrate this discrepancy we show the cumulative
size distributions
of isothermal, multi-thermal, and magnetic flare energies in Fig.~10,
where we sample an identical event list, which is the common subset
of the three energy forms and contains 171 events. In Fig.~10 we show 
a cumulative size distribution of these events, constructed with 
the inverse rank-order plot.  Note that the three different forms of
energy differ by an approximate amount of $(E_{magn}/E_{th,multi})
\approx 13$ and $(E_{th,multi}/E_{th,iso}) \approx 14$. 
It is therefore imperative to derive the same form of energy when
comparing the occurrence probabilities from the size distributions 
of different datasets.

\subsection{	Thermal/Magnetic Energy Ratios			}

One key result of this study is the thermal/magnetic energy ratio,
for which we found a range of $E_{th}/E_{diss} \approx 2\%-40\%$.
We consider this result to be a substantial improvement over previous
estimates, where isothermal instead of multi-thermal temperature
distributions were used
and no measurements of magnetically dissipated energies were available, 
resulting into a much lower estimate of the thermal energy content
in the order of $E_{th}/E_{diss} \approx 0.2\%-1\%$ (Emslie et al.~2012). 
The thermal energy is smaller than the magnetically dissipated energy
for essentially all events (Fig.~8a), while the few mavericks can be
explained by inaccurate energy measurements, either on the thermal
or magnetic part. This result is certainly consistent with most
magnetic reconnection models (where magnetic energy is converted
into acceleration of particles) and the thick-target model (where
the accelerated particles lose their energy by precipitation down
to the chromosphere and heat up the chromospheric plasma). The
amount of energy that goes into chromospheric and coronal plasma
heating may well be larger than the thermal energy measured here,
because we measured only the thermal energy content at the peak time
of the flare, while multiple heating phases may occur before and
after the flare peak. Even if we would add up all thermal energies
from every flare episode that shows a subpeak in the soft or hard
X-ray time profile, we would still underestimate the thermal energy
because (radiative and conductive) cooling processes are not considered in 
the calculation of the thermal energy content here. 
Thus, the multi-thermal energy content calculated here
represents only a lower limit of the heating energy that goes into
flare plasma heating during a flare. A complete calculation of the
multi-thermal flare energy would require a forward-fitting method
of the evolution of the heating rate $dE_h/dt$ that fits the observed
conductive $dE_{cond}/dt$ and radiative energy loss rate $dE_{rad}/dt$,
which is beyond the scope of this study, since this would require 
realistic geometric 3D models of flare loop arcades also. 

\section{               CONCLUSIONS                             }

As part of a global flare energetics study that encompasses all forms
of energies that are converted during solar flares (with or without CMEs)
we calculated the dissipated magnetic energy of 172 GOES M- and X-class 
events (in Paper I), and the multi-thermal energy at the peak time of 391
flare events (in this Paper II here). The catalog of these flare events 
is available online, see {\sl http://www.lmsal.com/$\sim$aschwand/RHESSI/ 
flare$\_$energetics.html}.  The major results of this study are:

\begin{enumerate}

\item{We computed the differential emission measure (DEM) distribution 
function of all 391 flares in time steps of $\Delta t=0.1$ hr using the
spatially-synthesized Gaussian DEM forward-fitting method, which yields
a detailed shape of the multi-thermal DEM distribution. This method is
found to be robust and converges as a function of the macro-pixel 
size to a unique DEM solution, subject to uncertainties in terms of the 
instrumental response function and subtracted background fluxes in the
order of $\approx 10\%$. The multi-thermal DEM function yields a 
significantly higher (typically by a factor of $\approx 14$, 
but comprehensive, (multi-)thermal energy than the isothermal 
energy estimated from the same data.}

\item{For the overlapping dataset of 171 flare events for which we could
calculate both the magnetically dissipated energies $E_{diss}$ and the
multi-thermal energies $E_{th}$, we find a 
ratio of $E_{th}/E_{diss} \approx 2\%-40\%$. This value is about
an order of magnitude higher than previous estimates, i.e., 
$E_{th}/E_{diss} \approx 0.2\%-1.0\%$, where isothermal energies from
GOES X-ray data rather than multi-thermal energies from EUV AIA data
were calculated, and a ratio of $E_{diss}/E_p
=30\%$ was assumed ad hoc (Emslie et al.~2012).}

\item{The computed thermal energies are consistent
with the RTV scaling law $E_{th,RTV} = 7.3 \times 10^{-10} \ 
T_p^3 L_p^2$, which applies to the energy balance between the heating
and (conductive and radiative) cooling rate at the turning point of
the flare peak time. In our analyzed dataset of M and X-class flares
we find thermal energies in the range of $E_{th}=10^{28.3}-10^{32.0}$ erg.
In comparison, the largest historical flare event has been reported to have
an energy of $E_{th} \approx 10^{33}$ erg, while the smallest coronal nanoflares
with a length scale of $L_{min} \gapprox 1$ Mm and coronal temperature of 
$T_e \gapprox 1$ MK are predicted to have values of $E_{th} \gapprox
10^{24}$ erg according to the RTV scaling law.}

\item{The size distributions of the spatial parameters display
a powerlaw tail with powerlaw slopes of 
$\alpha_L^{obs} = 3.3\pm0.3$ for the length scales, 
$\alpha_V^{obs} = 1.7\pm0.2$ for flare volumes, 
$\alpha_E^{obs} = 1.8\pm0.2$ for flare volumes, and are
consistent with the predictions of the fractal-diffusive
self-organized criticality model combined with the RTV scaling law
($\alpha_L = 3.0$; $\alpha_V = 1.67$; $\alpha_E = 1.88$).}

\end{enumerate}

After we have established the measurements of magnetically dissipated
flare energies (Paper I) and the multi-thermal energies (Paper II here),
we plan to measure the non-thermal energies (using RHESSI), the kinetic 
energies of CMEs (using AIA/SDO and STEREO), and the various radiative 
energies in gamma-rays, hard X-rays, soft X-rays, EUV, and bolometric 
luminosity in future studies. The ultimate goal is to quantify and
understand the energy partition in a comprehensive set of large
flare/CME events, and to identify the physical processes that are
consistent with the various flare energy measurements.  

\bigskip
\acknowledgements
We appreciate helpful and constructive comments from an anonymous
referee and from a number of participants of the RHESSI-13 workshop.
Part of the work was supported by
NASA contract NNG 04EA00C of the SDO/AIA instrument and
the NASA STEREO mission under NRL contract N00173-02-C-2035.

\clearpage

\section*{ APPENDIX A: Thermal Energy of a Multithermal DEM   }

Thermal energies of solar flares are generally estimated by the
expression for a homogeneous and isothermal plasma (Eq.~11),
$$
         E_{th} = 3 n_p k_B T_p V
                = 3 k_B T_p \sqrt{EM_p \ V} \ ,
	\eqno(A1)
$$
where $n_p = EM_p/V$ is the electron density, $T_p$ the electron
temperature, and $V$ the volume, measured at the peak time $t_p$
of a flare. The values $EM_p$ and $T_p$ are generally determined
from the peak in a DEM distribution function.

However, since the solar flare plasma is inhomogeneous and 
multi-thermal, we can calculate a more accurate expression for the
total thermal energy when imaging observations are available.
Ideally, such as in the case of an MHD simulation, the full 3D
distributions of temperatures $T_e(x,y,z)$ and electron densities
$n_e(x,y,z)$ are known, so that the most accurate expression for
thermal energies can be computed by volume integration (e.g.,
Testa et al.~2012),
$$
	E_{th} = \int \int \int 3 n_e(x,y,z) k_B T_e(x,y,z) \ dx \ dy\ dz \ .
	\eqno(A2)
$$
For numerical computations, we use a discretized 3D volume 
$(x_i, y_j, z_k)$ that is aligned in the z-direction with the line-of-sight, 
while images in different wavelengths have the 2D coordinate 
system $(x_i, y_i)$ with pixel size $\Delta x=\Delta y$. A DEM analysis
yields an inversion of a DEM distribution $DEM_{ij}(T)=DEM(T; x_i, y_j)$ 
in every pixel at location $(x_i,y_j)$. The column depth emission measure 
is defined by 
$$
	EM_{ij}=\int DEM_{ij}(T)\ dT = \int n_{ij}^2 \ dz = n_{ij}^2 \ L 
	\eqno(A3)
$$
which yields an average density $n_{ij}$ along the line-of-sight
column depth with length $L$ at each pixel position $(x_i, y_j$). 
We can then define a thermal energy $E_{th,ij}$ for each column depth $L=V^{1/3}$
by summing all contributions $EM_k$ from each temperature interval 
$\Delta T_k$ (Eq.~11),
$$
        E_{th,ij} 
	= \sum_k 3 k_B V^{1/2}\ T_{ijk} \ EM_{ij}^{1/2}
        = 3 k_B V^{1/2} \sum_k T_k \ \left[ DEM_{ij}(T_k)\ 
	\Delta T_k \right]^{1/2} \ .
	\eqno(A4)
$$
The total thermal energy in the computation box can then be obtained
by summing up the partial thermal energies $EM_{ij}$ from all pixels,
$$
        E_{th} = \sum_i \sum_j E_{th,ij} \Delta x^2  
        = 3 k_B V^{1/2} \sum_i \sum_j \sum_k \ T_k \left[ DEM_{ij}(T_k)\ 
	\Delta T_k \right]^{1/2} \Delta x^2
$$
$$
        = 3 k_B V^{1/2} \sum_k \ T_k \left[ \sum_i \sum_j DEM_{ij}(T_k)\ 
	\Delta T_k \right]^{1/2} \Delta x^2
        = 3 k_B V^{1/2} \sum_k \ T_k \left[ DEM(T_k)\ 
	\Delta T_k \right]^{1/2} \ ,
	\eqno(A5)
$$
where we replaced the partial DEM functions $DEM_{ij}(T_k)$ per column depths
by the total DEM function $DEM(T_k)$,
$$
	DEM(T_k) = \sum_i \sum_j DEM_{ij}(T_k) \Delta x^2\ ,
	\eqno(A6)
$$
which leads to the expression given in Eq.~(12). 

We compare now the thermal energy $E_{th}$ (Eq.~A5) computed in this way 
for a multi-thermal DEM distribution with the isothermal approximation
(Eq.~A1) by their ratio in Fig.~11, given for a set of thermal widths
$w_T=0.1, 0.2, ..., 1.0$ in the single-Gaussian DEM function
(Eq.~1) that is used for DEM modeling in each pixel. For small values,
say $w_T=0.1$, the DEM distributions are almost isothermal, and
thus the approximation (Eq.~A1) is appropriate and we obtain a ratio
near unity $q_{iso}=(E_{iso}/E_{multi}) \gapprox 1$). For broader multi-thermal
DEM functions, the ratio increases systematically, up to a factor of
$q_{iso} \lapprox 30$. At higher temperatures, the ratio decreases because
the temperature range between the peak of the DEM and the upper limit
(here at $T=30$ MK) becomes increasingly smaller and thus has less
weight in the asymmetric $T$-weighting of the thermal energy contributions. 
Observed DEM peaks have typically a logarithmic temperature half width 
of $w_T \approx 0.5$ (see Fig.~4 for examples), and thus the multi-thermal 
energy ratios vary by a factor of $q_{iso} \approx 2 - 8$ for flare peak 
temperatures in the range of $T_p \approx 1-10$ MK (Fig.~11). 
Since the observed DEM distributions are generally multi-peaked, 
the ratios tend to be higher than estimated from single-Gaussian DEMs 
as shown in Fig.~(11).


\clearpage

\begin{deluxetable}{rrrrrrrrrr}
\tabletypesize{\footnotesize}
\tablecaption{Thermal energy parameters of 28  
X-class flare events.}
\tablewidth{0pt}
\tablehead{
\colhead{\#}&
\colhead{Flare}&
\colhead{GOES}&
\colhead{Helio-}&
\colhead{Length}&
\colhead{Peak}&
\colhead{EM-weighted}&
\colhead{Electron}&
\colhead{Emission}&
\colhead{Thermal}\\
\colhead{}&
\colhead{start time}&
\colhead{class}&
\colhead{graphic}&
\colhead{scale}&
\colhead{temperature}&
\colhead{temperature}&
\colhead{density}&
\colhead{measure}&
\colhead{energy}\\
\colhead{}&
\colhead{}&
\colhead{}&
\colhead{position}&
\colhead{$L$}&
\colhead{$T_p$}&
\colhead{$T_w$}&
\colhead{$log(n_e)$}&
\colhead{$log(EM)$}&
\colhead{$E_{th}$}\\
\colhead{}&
\colhead{}&
\colhead{}&
\colhead{}&
\colhead{(Mm)}&
\colhead{(MK)}&
\colhead{(MK)}&
\colhead{(cm$^{-3})$}&
\colhead{(cm$^{-3})$}&
\colhead{$(10^{30}$ erg)}\\}
\startdata
  12 & 20110215 0144 & X2.2 & S21W12 &   28.4 &   15.9 &   27.9 &   10.8 &   49.9 &   82.2 \\ 
  37 & 20110309 2313 & X1.5 & N10W11 &   34.8 &    5.6 &   22.5 &   10.6 &   49.7 &   84.7 \\ 
  61 & 20110809 0748 & X6.9 & N14W69 &   28.9 &   15.9 &   28.4 &   10.9 &   50.2 &  128.9 \\ 
  66 & 20110906 2212 & X2.1 & N16W15 &   24.5 &   15.9 &   25.4 &   10.8 &   49.7 &   52.2 \\ 
  67 & 20110907 2232 & X1.8 & N16W30 &   37.4 &   28.2 &   28.9 &   10.6 &   50.0 &  140.6 \\ 
 107 & 20111103 2016 & X1.9 & N21E64 &   26.2 &   25.1 &   33.2 &   10.8 &   49.8 &   76.4 \\ 
 132 & 20120127 1737 & X1.7 & N33W85 &   46.0 &    6.3 &   14.8 &   10.4 &   49.8 &  107.3 \\ 
 136 & 20120305 0230 & X1.1 & N19E58 &   29.7 &   14.1 &   33.9 &   10.7 &   49.8 &   92.9 \\ 
 147 & 20120307 0002 & X5.4 & N18E31 &   44.9 &   14.1 &   21.8 &   10.6 &   50.3 &  208.4 \\ 
 148 & 20120307 0105 & X1.3 & N18E29 &   36.0 &    4.0 &   22.7 &   10.5 &   49.7 &   89.8 \\ 
 209 & 20120706 2301 & X1.1 & S13W59 &   20.4 &    5.6 &   28.5 &   10.8 &   49.5 &   33.2 \\ 
 220 & 20120712 1537 & X1.4 & S15W03 &   36.3 &    6.3 &   24.2 &   10.6 &   49.8 &  105.3 \\ 
 248 & 20121023 0313 & X1.8 & S13E58 &   10.4 &   15.9 &   34.1 &   11.1 &   49.4 &   11.5 \\ 
 286 & 20130513 1548 & X2.8 & N08E89 &   23.6 &   17.8 &   33.3 &   10.9 &   49.9 &   70.9 \\ 
 287 & 20130514 0000 & X3.2 & N08E77 &   29.9 &   15.9 &   28.5 &   10.8 &   50.1 &  109.5 \\ 
 288 & 20130515 0125 & X1.2 & N10E68 &   22.6 &    5.6 &   27.1 &   10.7 &   49.5 &   40.8 \\ 
 318 & 20131025 0753 & X1.7 & S08E73 &   11.4 &   28.2 &   29.1 &   11.1 &   49.3 &   10.9 \\ 
 320 & 20131025 1451 & X2.1 & S06E69 &   17.0 &    4.0 &   30.5 &   10.9 &   49.4 &   24.6 \\ 
 330 & 20131028 0141 & X1.0 & N05W72 &   15.9 &   15.9 &   32.0 &   10.9 &   49.5 &   23.3 \\ 
 337 & 20131029 2142 & X2.3 & N05W87 &   23.9 &   22.4 &   30.3 &   10.8 &   49.8 &   53.7 \\ 
 344 & 20131105 2207 & X3.3 & S08E44 &   12.0 &   25.1 &   32.8 &   11.1 &   49.4 &   14.7 \\ 
 349 & 20131108 0420 & X1.1 & S11E11 &   20.8 &   25.1 &   30.2 &   10.9 &   49.8 &   43.0 \\ 
 351 & 20131110 0508 & X1.1 & S13W13 &   22.0 &   17.8 &   33.3 &   10.9 &   49.8 &   54.5 \\ 
 358 & 20131119 1014 & X1.0 & S13W69 &   18.3 &   20.0 &   30.7 &   10.9 &   49.6 &   31.4 \\ 
 384 & 20140107 1804 & X1.2 & S12E08 &    3.0 &    1.8 &   25.1 &   11.4 &   48.2 &    0.4 \\ 
\enddata
\end{deluxetable}


\begin{deluxetable}{rrrrrrrrrr}
\tabletypesize{\footnotesize}
\tablecaption{Thermal energy parameters of 391 M and X-class flare events.
The full list is available from a machine-readable file, from which only the
first 10 entries are listed here.}
\tablewidth{0pt}
\tablehead{
\colhead{\#}&
\colhead{Flare}&
\colhead{GOES}&
\colhead{Helio-}&
\colhead{Length}&
\colhead{Peak}&
\colhead{EM-weighted}&
\colhead{Electron}&
\colhead{Emission}&
\colhead{Thermal}\\
\colhead{}&
\colhead{start time}&
\colhead{class}&
\colhead{graphic}&
\colhead{scale}&
\colhead{temperature}&
\colhead{temperature}&
\colhead{density}&
\colhead{measure}&
\colhead{energy}\\
\colhead{}&
\colhead{}&
\colhead{}&
\colhead{position}&
\colhead{$L$}&
\colhead{$T_p$}&
\colhead{$T_w$}&
\colhead{$log(n_e)$}&
\colhead{$log(EM)$}&
\colhead{$E_{th}$}\\
\colhead{}&
\colhead{}&
\colhead{}&
\colhead{}&
\colhead{(Mm)}&
\colhead{(MK)}&
\colhead{(MK)}&
\colhead{(cm$^{-3})$}&
\colhead{(cm$^{-3})$}&
\colhead{$(10^{30}$ erg)}\\}
\startdata
   1 & 20100612 0030 & M2.0 & N23W47 &   13.2 &    6.3 &   18.6 &   10.8 &   48.9 &    7.0 \\ 
   2 & 20100613 0530 & M1.0 & S24W82 &   12.2 &    7.1 &   15.8 &   10.7 &   48.6 &    4.1 \\ 
   3 & 20100807 1755 & M1.0 & N13E34 &   25.1 &    4.0 &    6.2 &   10.4 &   49.1 &    8.2 \\ 
   4 & 20101016 1907 & M2.9 & S18W26 &   15.1 &   14.1 &   29.9 &   10.9 &   49.4 &   19.2 \\ 
   5 & 20101104 2330 & M1.6 & S20E85 &   13.8 &   10.0 &   24.9 &   10.8 &   49.0 &   10.2 \\ 
   6 & 20101105 1243 & M1.0 & S20E75 &   13.3 &    6.3 &   19.8 &   10.8 &   48.9 &    7.7 \\ 
   7 & 20101106 1527 & M5.4 & S20E58 &   20.3 &    4.0 &   20.9 &   10.7 &   49.4 &   24.6 \\ 
   9 & 20110209 0123 & M1.9 & N16W70 &    8.2 &    3.2 &   28.5 &   11.0 &   48.7 &    3.4 \\ 
  10 & 20110213 1728 & M6.6 & S21E04 &   15.9 &   14.1 &   23.9 &   10.9 &   49.5 &   20.8 \\ 
 ... & ...           & ...  & ...    &   ...  &   ...  &   ...  &   ...  &   ...  &   ...  \\
\enddata
\end{deluxetable}
\clearpage

\begin{deluxetable}{lrrrrrr}
\tablecaption{Parameter ranges of physical parameters determined from 4 different
datasets of large flares: $^{1})$ Pallavicini et al.~(1977);
                          $^{2})$ Emslie et al.~(2012);
			  $^{3})$ Aschwanden and Shimizu (2013);
			  $^{4})$ This study: Isothermal energy;
			  $^{5})$ This study: Multi-thermal energy.}
\tablewidth{0pt}
\tablehead{
\colhead{Instrument}&
\colhead{Number}&
\colhead{Spatial}&
\colhead{Electron}&
\colhead{Electron}&
\colhead{Emission}&
\colhead{Thermal}\\
\colhead{}&
\colhead{of events}&
\colhead{scale}&
\colhead{temperature}&
\colhead{density}&
\colhead{measure}&
\colhead{energy}\\
\colhead{}&
\colhead{$n$}&
\colhead{$log(L)$}&
\colhead{$log(T_w)$}&
\colhead{$log(n_e)$}&
\colhead{$log(EM_p)$}&
\colhead{$log(E_{th})$}\\
\colhead{}&
\colhead{}&
\colhead{[Mm]}&
\colhead{[MK]}&
\colhead{[cm$^{-3}$]}&
\colhead{[cm$^{-3}$]}&
\colhead{[erg]}}
\startdata
Skylab/S-054$^{1}$    &  31 & $8.7-9.7$  & $6.8-7.1$   & $ 9.9-11.3$   & $40.1-49.3$  & $28.6-31.0$  \\
RHESSI$^{2}$          &  32 &            &             &               &              & $30.0-31.3$  \\
AIA/SDO$^{3}$	      & 155 & $8.6-9.8$  & $6.1-7.3$   & $ 9.6-11.9$  & $47.0-50.6$   & $28.3-32.0$  \\
AIA/SDO$^{4}$	      & 391 & $8.2-9.7$  & $5.7-7.4$   & $10.3-11.8$  & $47.3-50.3$   & $28.3-31.7$  \\
AIA/SDO$^{5}$	      & 391 & $8.2-9.7$  & $6.8-7.6$   & $10.3-11.8$  & $47.3-50.3$   & $28.3-32.0$  \\
\enddata
\end{deluxetable}
\rm
\clearpage


\begin{figure}
\plotone{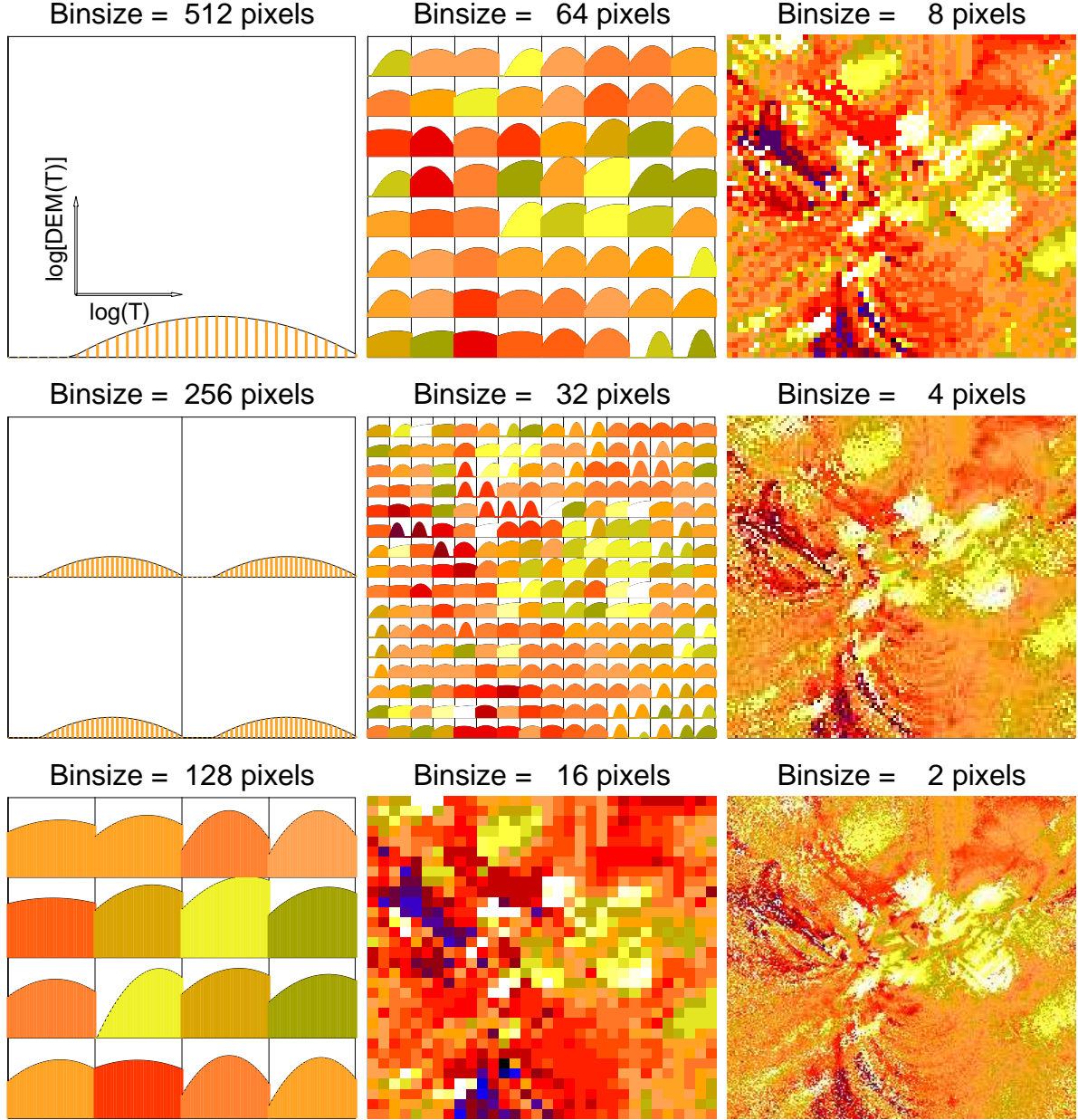}
\caption{The spatial synthesis DEM method is visualized by
single-Gaussian DEM fits in a grid $[x_i, y_j]$, $i=1,...,n_{bin},
j=1,...,n_{bin}$ of spatial positions with macropixels of bin size 
$\Delta x=512/n_{bin}=512, 256, ..., 2$. Each macropixel shows a
Gaussian DEM fit to the 6 coronal AIA wavelengths, covering a
temperature range of $log(T)=5.8-7.45$ and emission measure range of
$DEM(T)=10^{47}-10^{57}$ cm$^{-5}$ K$^{-1}$. The colorscale is
proportional to the logarithmic DEM peak temperature $T_p$, with
blue at $T_p=10^{5.8}$ K and white at $T=10^{7.45}$ K. 
The data are obtained from event \#12, a GOES X2.2-class flare 
observed with AIA on 2011 February 15, 01:40 UT.}
\end{figure}
\clearpage

\begin{figure}
\plotone{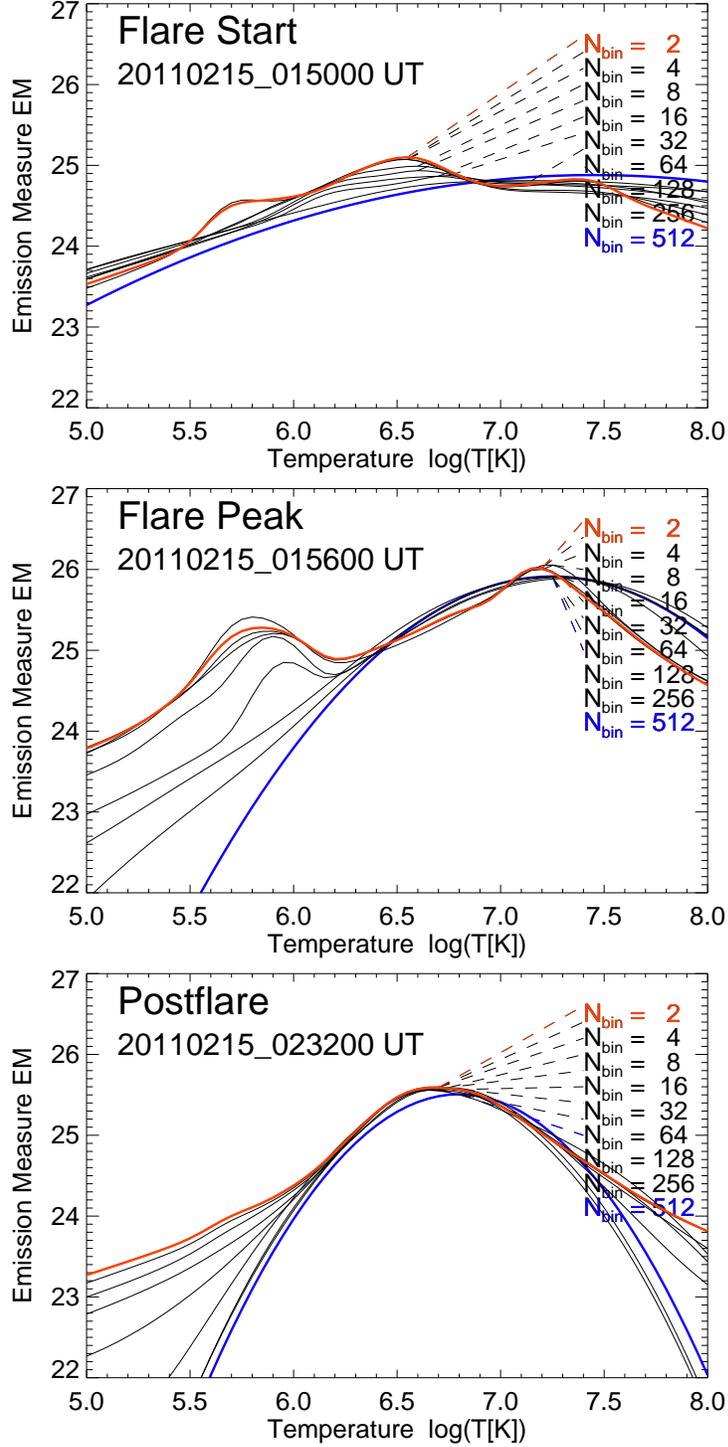}
\caption{Spatial convergence of DEM distribution functions (from case shown
in Fig.~1), as a function of the macro-pixel size, from $N_{bin}$=512 (blue) 
to $N_{bin}=2$ (red), using the spatial synthesis DEM method, shown for three
time steps during the 2011 Feb 15 flare, at flare start 01:50 UT (top panel),
at flare peak 01:56 UT (middle panel), and in the postflare phase at 02:32 UT
(bottom panel).}
\end{figure}
\clearpage

\begin{figure}
\plotone{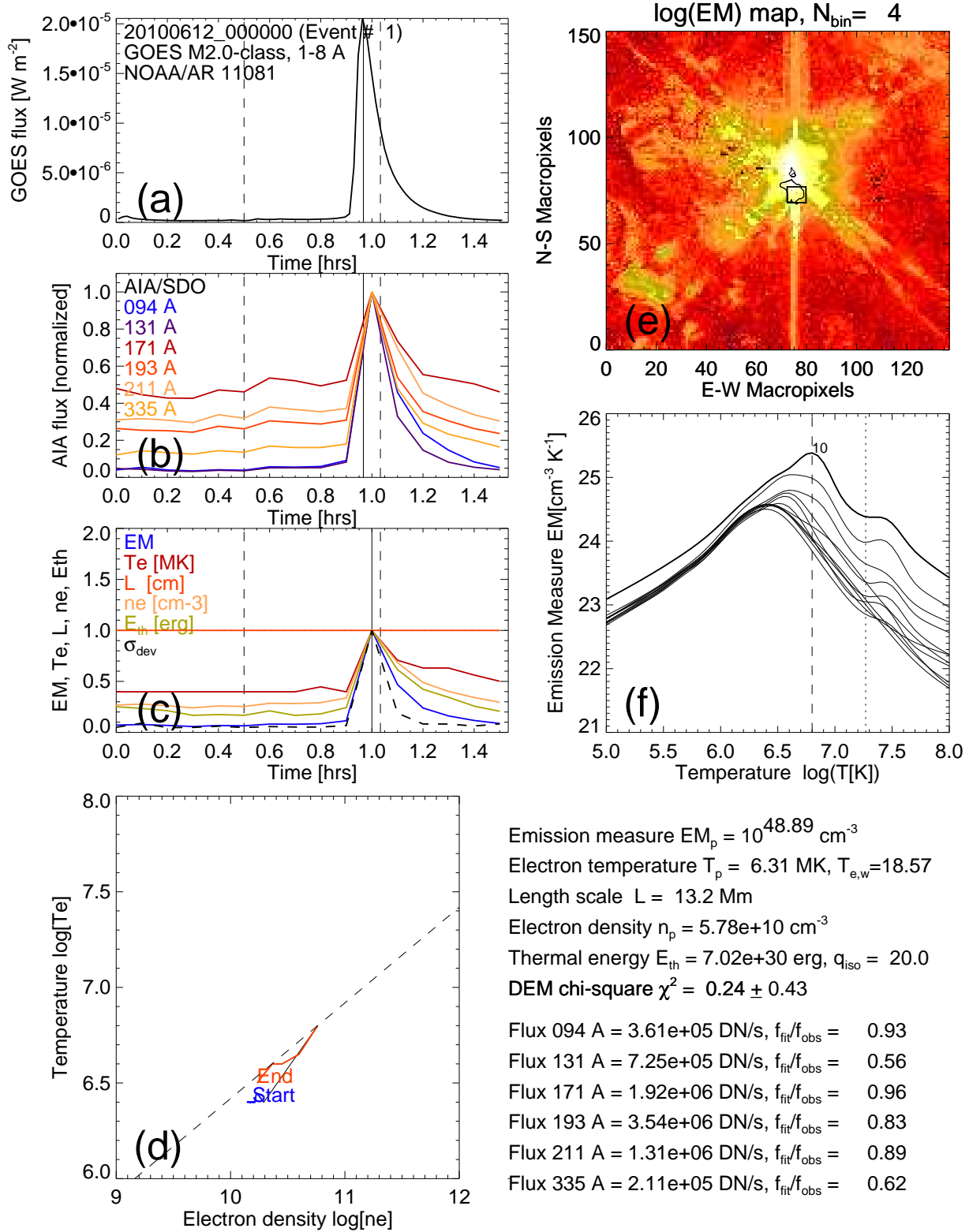}
\caption{A summary of the DEM modeling of event \#1, a GOES M2.0-class
flare observed with AIA on 2010 June 12, 00:00 UT: (a) GOES
1-8 \ang\ light curve with flare peak time (solid vertical line),
start and end times (dashed vertical lines; (b) the background-subtracted
light curves in the 6 coronal EUV channels from AIA/SDO (normalized
to unity); (c) the evolution of physical parameters; (d) a $T_e-n_e$ 
phase diagram (with the RTV equilibrium indicated by a
dashed line); (e) emission measure map $EM(x,y)$ at the flare peak;
(f) the spatial-synthesized DEM functions for all 14
time steps, with the emission measure maximum at time step 10; 
and the values of physical parameters
at the flare peak time (bottom right).}
\end{figure}
\clearpage

\begin{figure}
\plotone{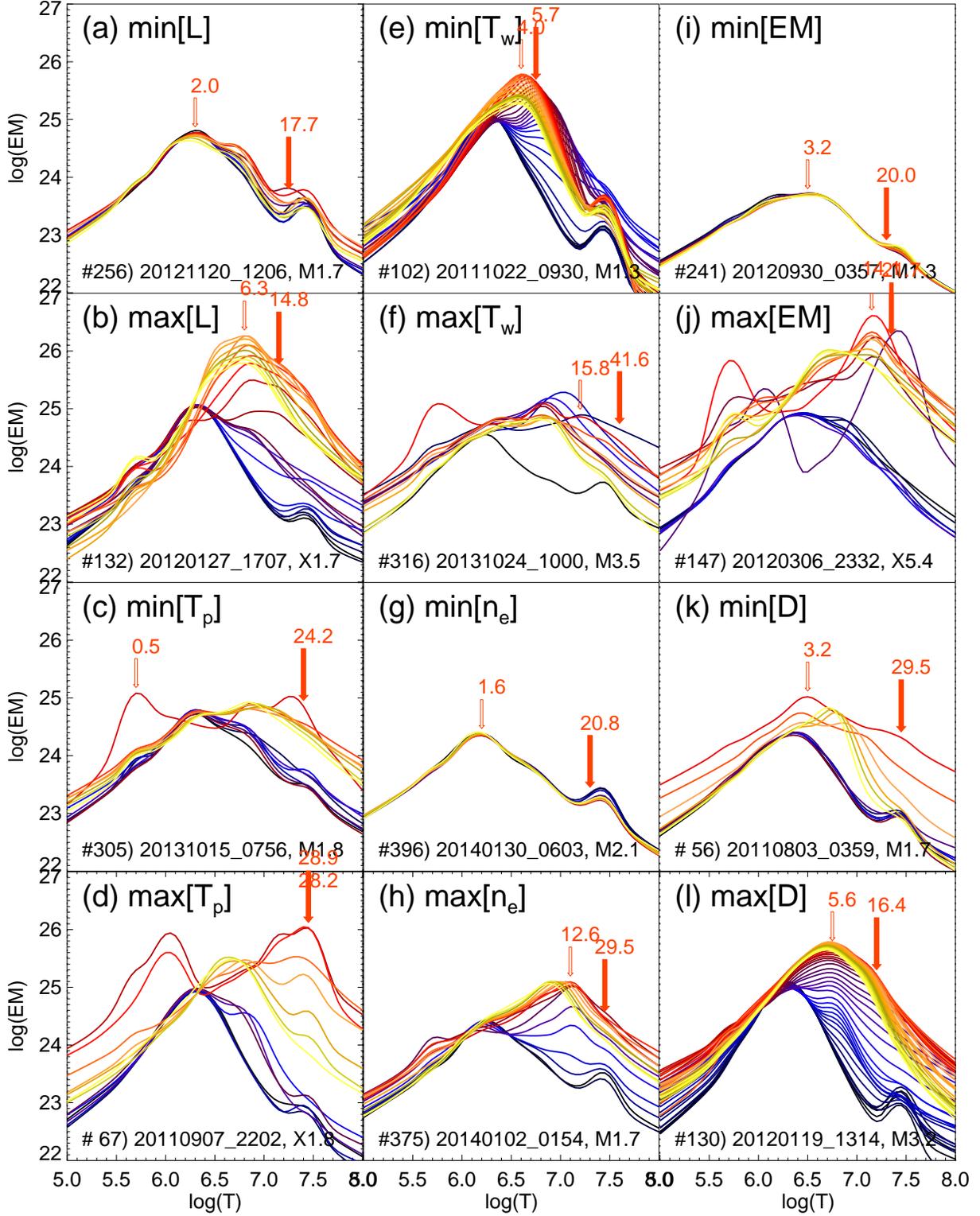}
\caption{Differential emission measure distributions $DEM(T)$ of 12 
extremal flares, calculated with the spatial synthesis DEM 
method, are shown in evolutionary time steps of $dt=0.1$ hr. The color
scale indicates the transition from preflare (blue) to flare peak time
(red) and postflare phase (yellow). The DEM peak temperatures $T_p$
at the peak time $t_p$ of the flare (red arrow) and the emission 
measure-weighted temperatures $t_w$ (red solid arrow) are indicated
in units of MK.}
\end{figure}
\clearpage

\begin{figure}
\plotone{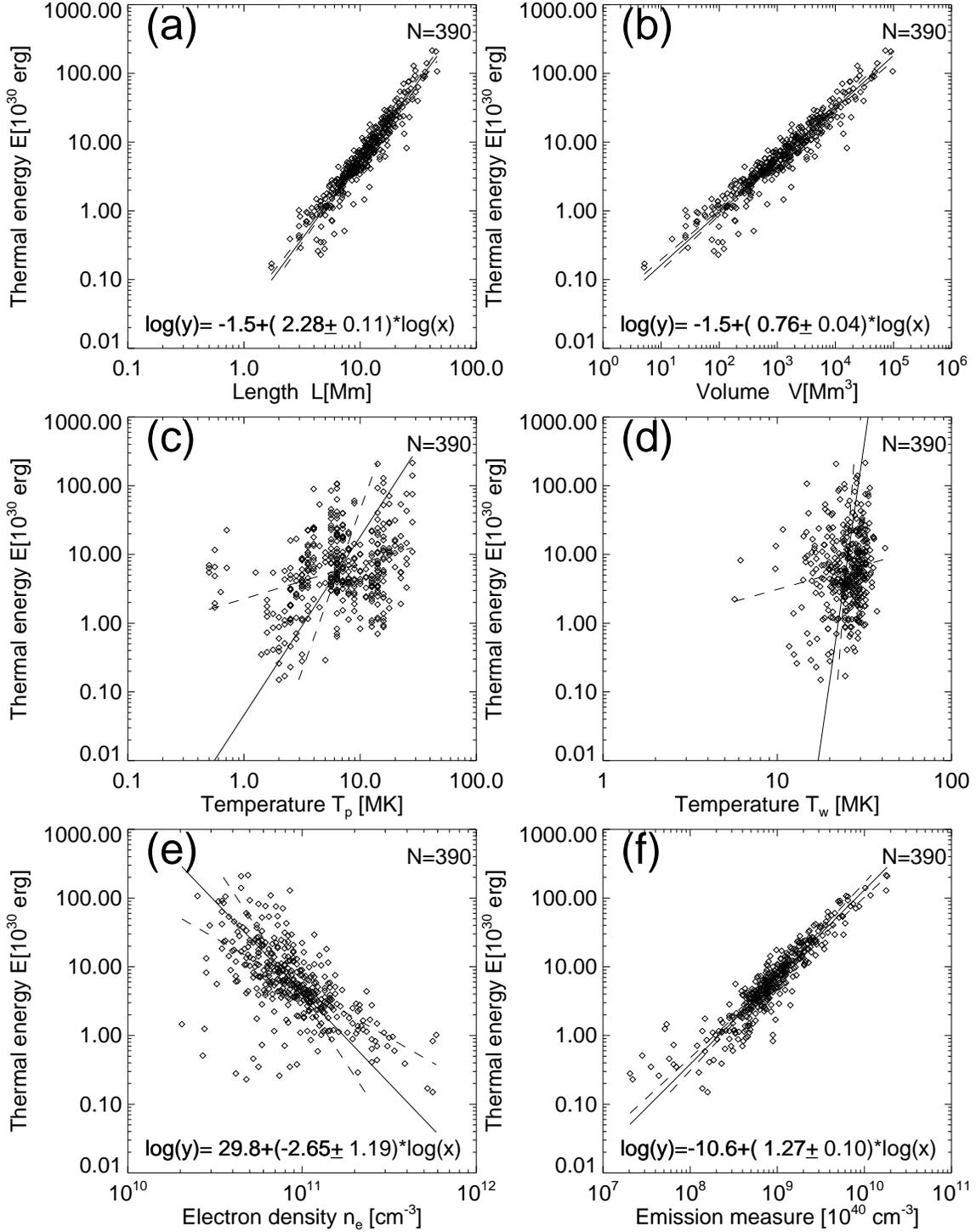}
\caption{Scatter plot ot the thermal energy $E_{th}$ as a function of
physical parameters $L, V, T_p, T_w, n_e, EM_p$ of the analyzed 391 
M and X-class flares.  Linear regression fits (solid lines) are 
indicated with 1-$\sigma$ uncertainties corresponding to the 67\%
confidence level (dashed lines).}
\end{figure}

\begin{figure}
\plotone{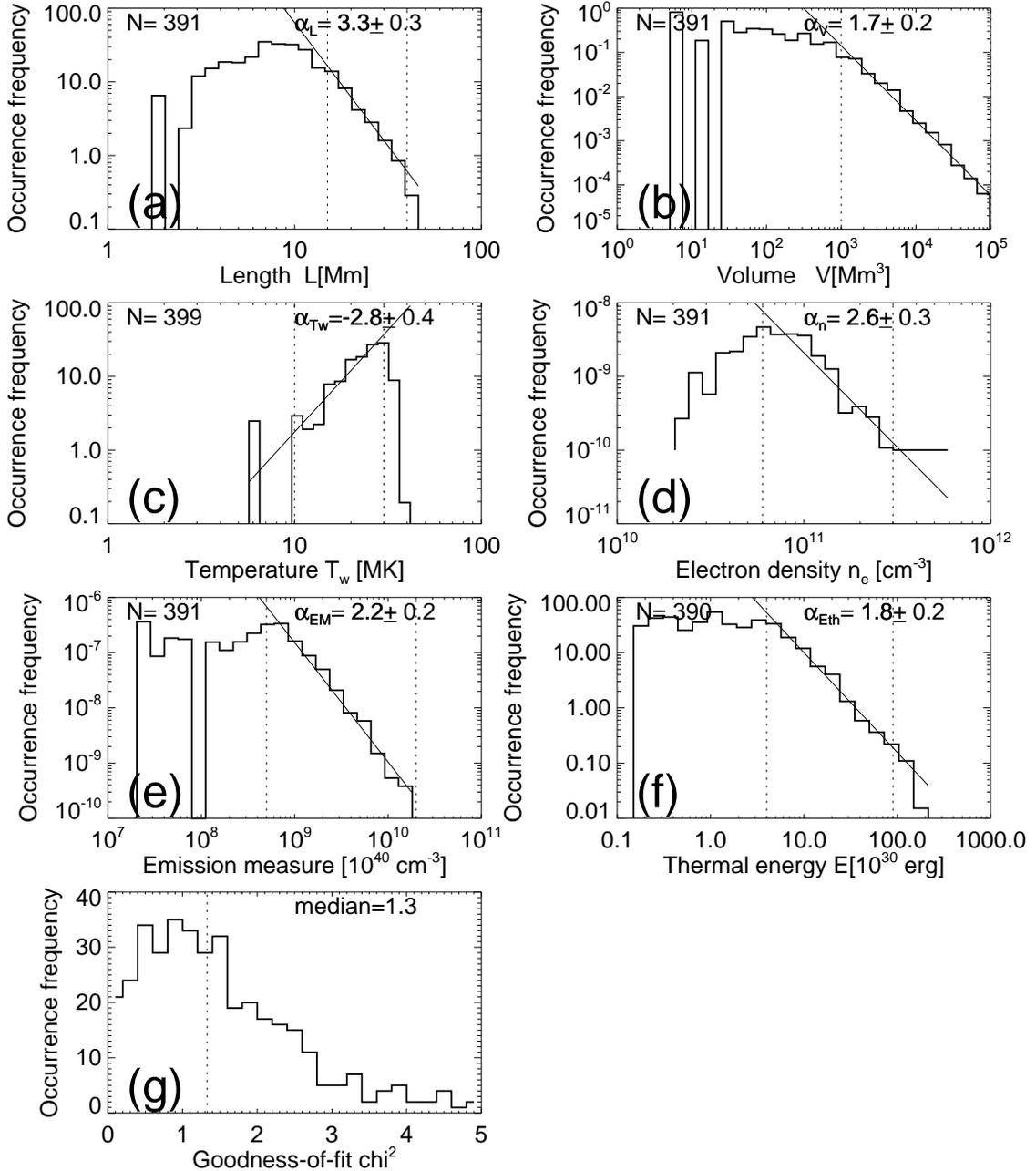}
\caption{Size distributions of the physical parameters
$L, V, T_w, n_e, EM$ and $E_{th}$ for the 391 analyzed M and X-class
flares. A powerlaw function is fitted in the range indicated with
dotted vertical lines. The reduced $\chi^2$ distribution is 
characterized with a normal distribution.}
\end{figure}

\begin{figure}
\plotone{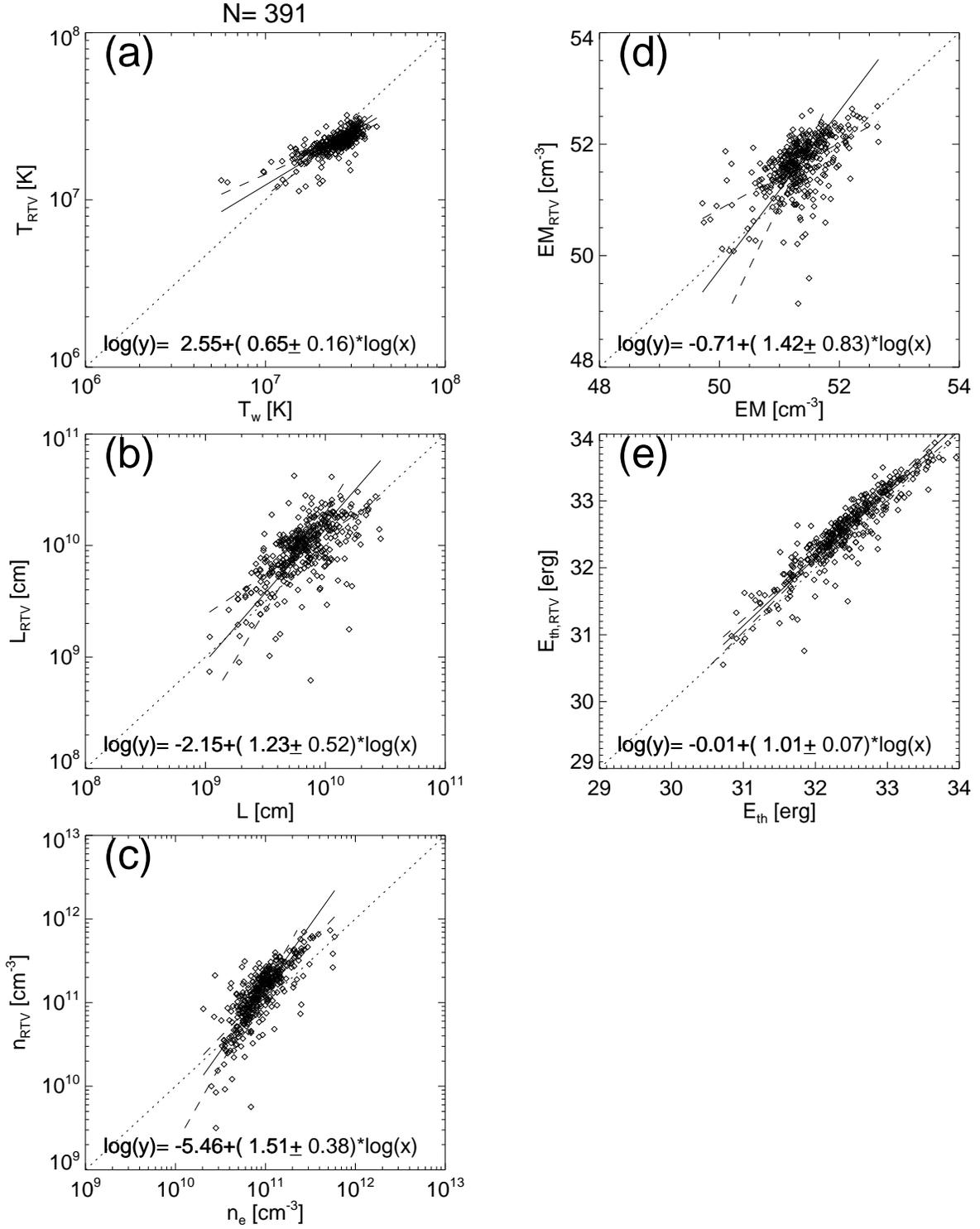}
\caption{Observed (x-axis) and predicted physical parameters (y-axis)
based on the Rosner-Tucker-Vaiana model. Linear regression fits (solid
lines) and uncertainties (dashed lines) are indicated,
along with the line for equivalence (dotted line).}
\end{figure}

\begin{figure}
\plotone{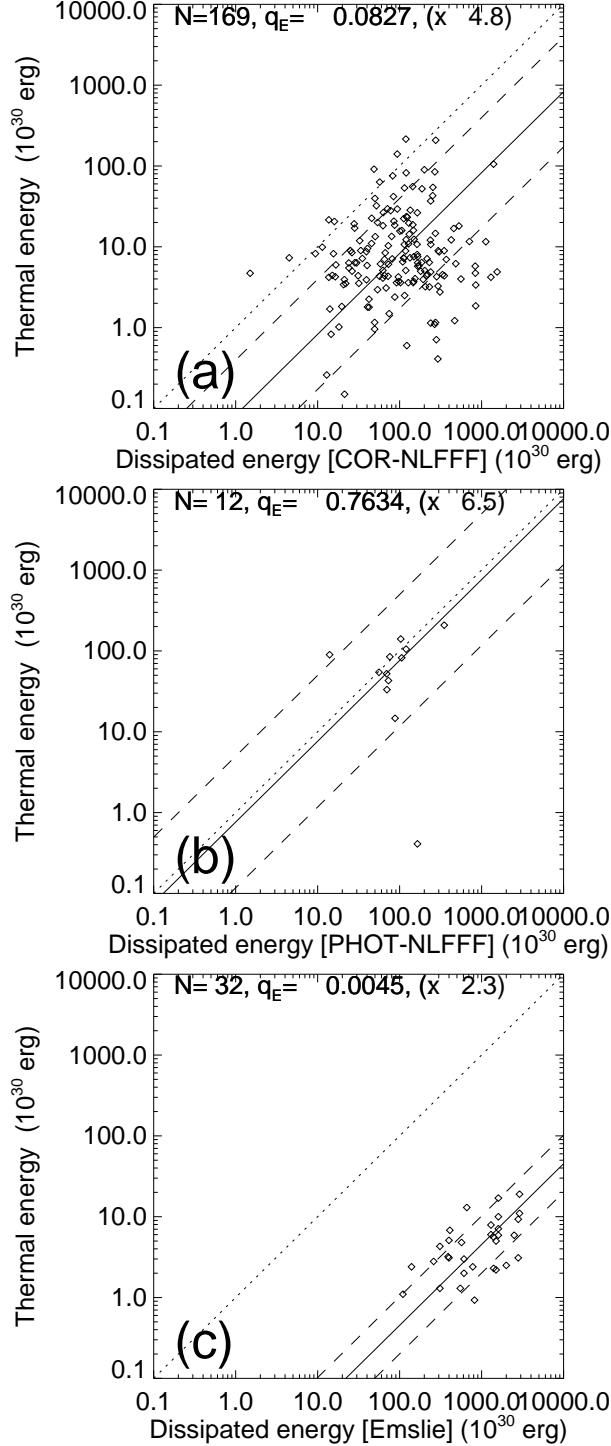}
\caption{Scatterplot of thermal energies $E_{th}$ versus magnetically
dissipated energies $E_{diss}$: (a) The 172 M and X-class flares 
from which the magnetically dissipated energy was determined in Paper I
(Aschwanden et al.~2014) with the COR-NLFFF method; (b) 12 events with
magnetic energies calculated with the PHOT-NLFFF method; (c)
32 large eruptive flares from Emslie et al.~(2012).  The mean ratio 
$q_E$ (solid line) and standard deviations (dashed lines, expressed
by a multiplication factor $\times$) are indicated, 
along with the line for equivalence (dotted line).}
\end{figure}

\begin{figure}
\plotone{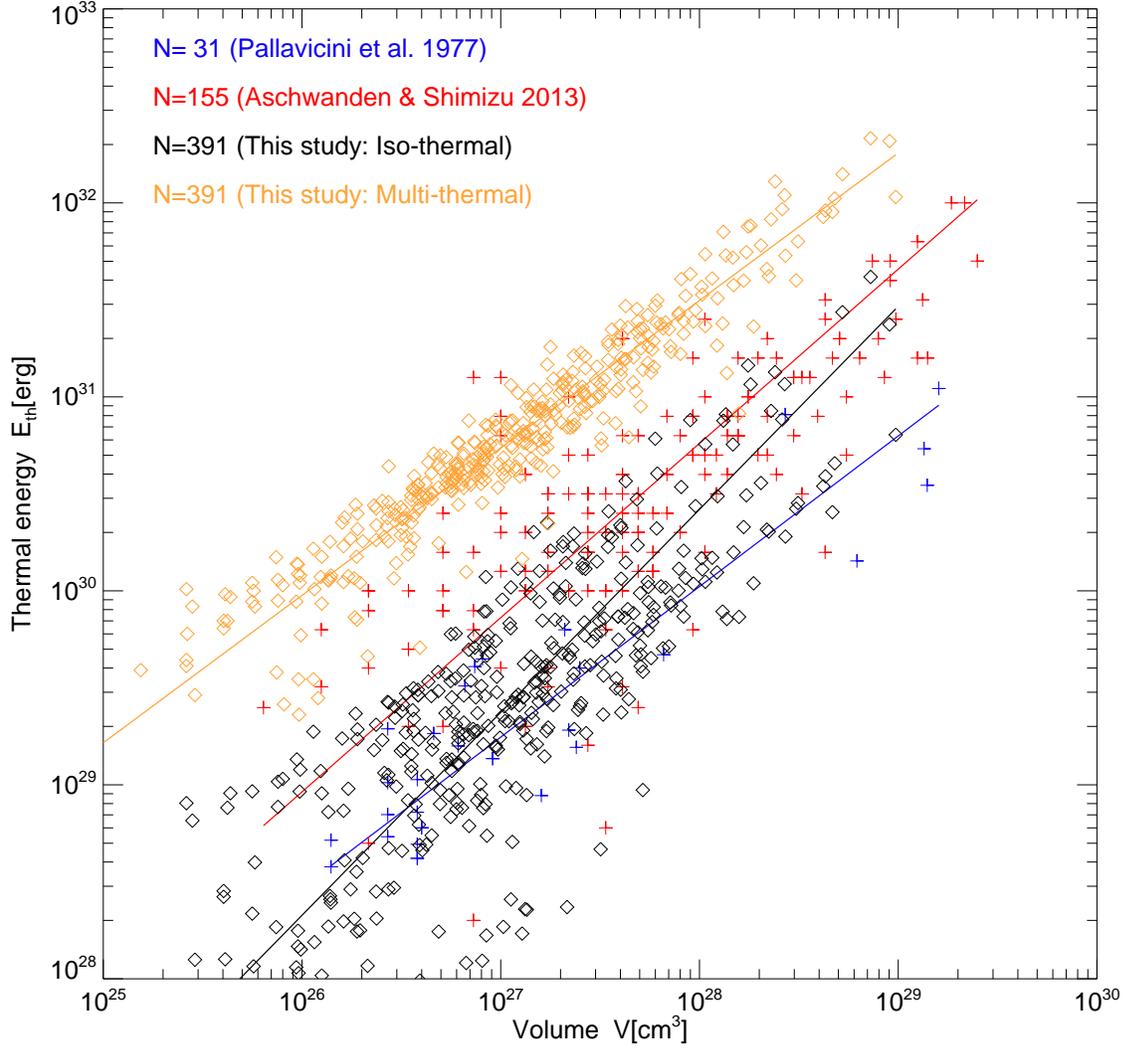}
\caption{Comparison of thermal energies as a function of the flare
volume size for 4 sets of measurements: Pallavicini et al.~(1977)
(blue crosses), Aschwanden and Shimizu (2013) (red crosses), 
isothermal energy in this study (black diamonds), 
and multi-thermal energy in this study (orange diamonds).
Note that the multi-thermal energies are about an order of magnitude
higher than the isothermal energies.}
\end{figure}

\begin{figure}
\plotone{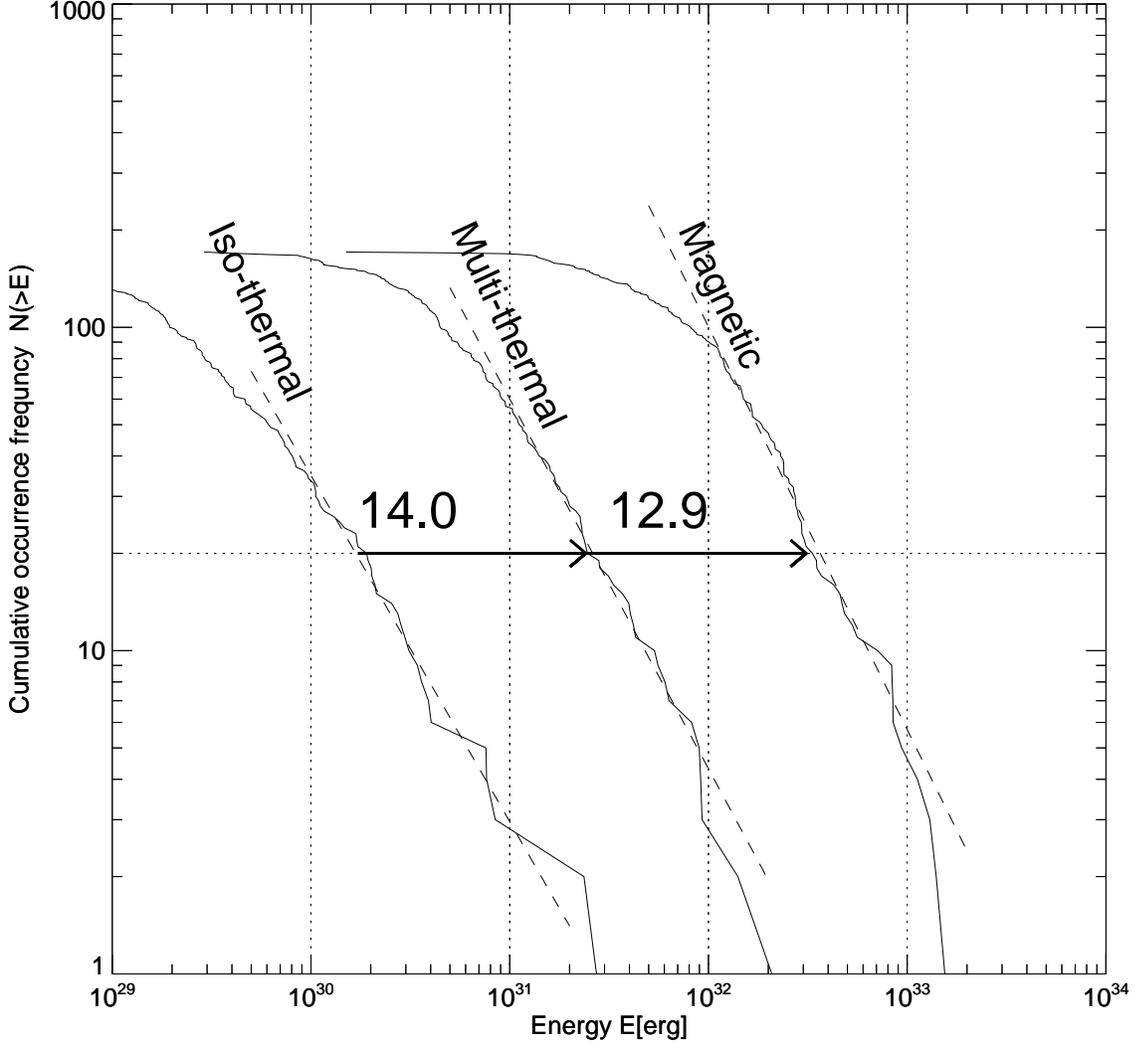}
\caption{Cumulative occurrence frequency distributions of isothermal,
multi-thermal, and dissipated magnetic energies in 171 M- and X-class
flares. Note that the energy values have a characteristic ratio of
$E_{th,multi}/E_{th,iso} \approx 14$ and 
$E_{magn}/E_{th,multi} \approx 12.9$.}
\end{figure}

\begin{figure}
\plotone{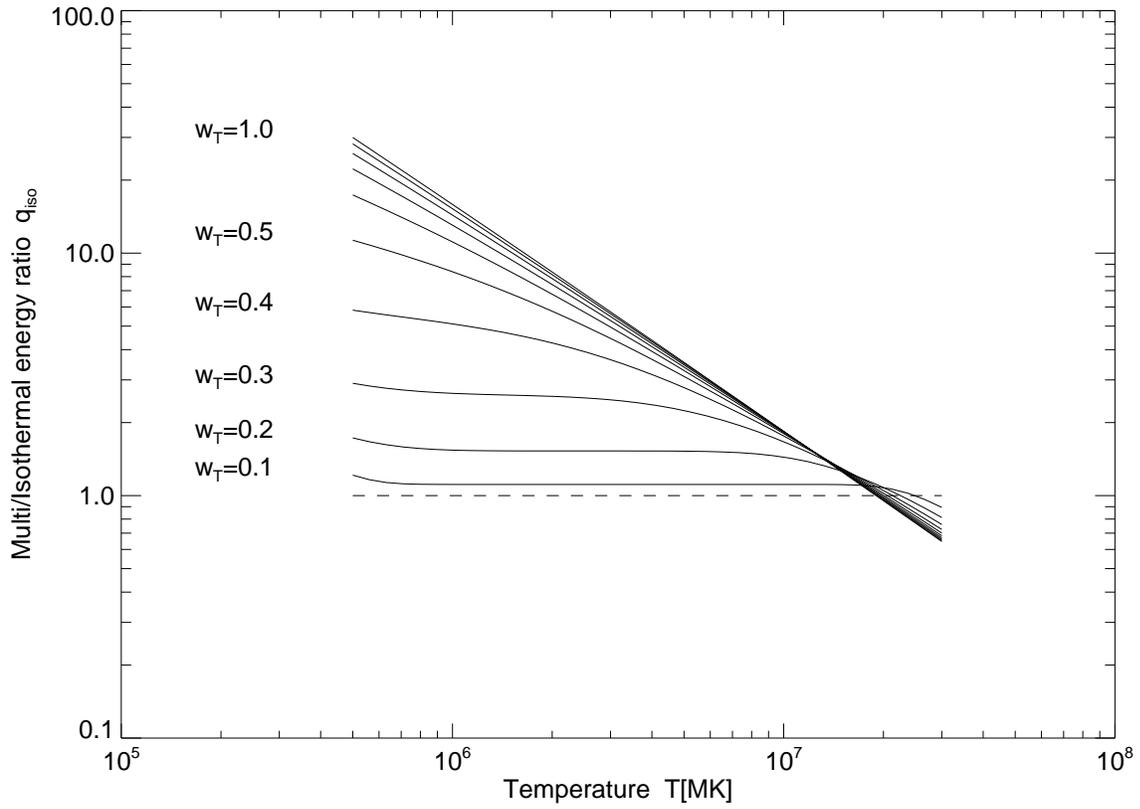}
\caption{The thermal energy ratio $q_{iso}=E_{th,multi}/E_{th,iso}$ 
is computed for narrow and broad single-Gaussian DEM distributions, 
with logarithmic temperature half widths (Eq.~1) in the range of 
$w_T \approx 0.1-1.0$.}
\end{figure}
\clearpage

\end{document}